\pdfoutput=1
\documentclass[iop]{emulateapj}
\usepackage{apjfonts}
\usepackage{amssymb,epsfig,mathptmx}

\shortauthors{Mao et al.}
\shorttitle{INFLUENCES OF DUST ATTENUATION AND STELLAR POPULATION AGE}

\begin{document}

\title{CHARACTERIZING ULTRAVIOLET AND INFRARED OBSERVATIONAL PROPERTIES 
FOR GALAXIES. I. INFLUENCES OF DUST ATTENUATION AND STELLAR POPULATION AGE}

\author{
YE-WEI MAO\altaffilmark{1,2}, 
ROBERT C. KENNICUTT, JR.\altaffilmark{3,4}, 
CAI-NA HAO\altaffilmark{5}, 
XU KONG\altaffilmark{1,2}, AND XU ZHOU\altaffilmark{6}}

\altaffiltext{1}{Center for Astrophysics, University of Science and
Technology of China, Hefei 230026, China; owen81@mail.ustc.edu.cn, xkong@ustc.edu.cn} 
\altaffiltext{2}{Key Laboratory for Research in Galaxies and Cosmology, 
USTC, CAS, Hefei 230026, China} 
\altaffiltext{3}{Institute of Astronomy, University of Cambridge, 
Madingley Road, Cambridge CB3 0HA, UK}
\altaffiltext{4}{Steward Observatory, University of Arizona, Tucson,
AZ 85721, USA} 
\altaffiltext{5}{Tianjin Astrophysics Center, Tianjin
Normal University, Tianjin 300387, China} 
\altaffiltext{6}{National
Astronomical Observatories, Chinese Academy of Sciences, Beijing
100012, China}

\begin{abstract}
The correlation between infrared-to-ultraviolet luminosity ratio and
ultraviolet color (or ultraviolet spectral slope), i.e., the IRX-UV
(or IRX-$\beta$) relation, found in studies of starburst galaxies is
a prevalent recipe for correcting extragalactic dust attenuation.
Considerable dispersion in this relation discovered for normal
galaxies, however, complicates its usability. In order to
investigate the cause of the dispersion and to have a better
understanding of the nature of the IRX-UV relation, in this paper,
we select five nearby spiral galaxies, and perform spatially
resolved studies on each of the galaxies, with a 
combination of ultraviolet and infrared imaging data. We measure all
positions within each galaxy and divide the extracted regions into
young and evolved stellar populations. By means of this approach, we
attempt to discover separate effects of dust attenuation and stellar
population age on the IRX-UV relation for individual galaxies. In
this work, in addition to dust attenuation, stellar population age
is interpreted to be another parameter in the IRX-UV function, and
the diversity of star formation histories is suggested to disperse
the age effects. At the same time, strong evidence shows the
necessity of more parameters in the interpretation of observational
data, such as variations in attenuation/extinction law. Fractional
contributions of different components to the integrated
luminosities of the galaxies suggest that the integrated
measurements of these galaxies, which comprise different populations would
weaken the effect of the age parameter on IRX-UV diagrams. The
dependance of the IRX-UV relation on luminosity and radial distance
in galaxies presents weak trends, which offers an implication of
selective effects. The two-dimensional maps of the UV color and the
infrared-to-ultraviolet ratio are displayed and show a disparity in
the spatial distributions between the two parameters in galaxies,
which offers a spatial interpretation of the scatter in the IRX-UV
relation.
\end{abstract}


\section{INTRODUCTION}

Obscuration of starlight by interstellar dust grains is a serious
obstacle which hampers our ability to directly derive stellar population properties of
galaxies from observed radiative information. Correction
for dust attenuation is crucial for determining galactic properties
such as star formation history (SFH) in current extragalactic
astronomy. Calibration of observational indicators to intrinsic
parameters of galaxies, for instance from observed UV-optical flux
to star formation rate (SFR), is in tight connection with the fact
of proper compensation made for dust attenuation
\citep[e.g.,][]{2004A&A...425..417K, 2009ApJ...703.1672K,
2009ApJ...706..599L, 2011ApJ...741..124H}.

Dust grains absorb stellar emission with the wavelength coverage
from far-ultraviolet to near-infrared, and re-emit the energy as
mid- and far-infrared thermal radiation. According to the energy
balance theory, the luminosity ratio of total infrared (IR) to
ultraviolet \citep[UV; so-called IR excess or IRX, initially
introduced in][]{1974A&A....32..269M} has been invented and
considered as a reliable indicator of dust attenuation
\citep{1992A&A...264..444B, 1995A&A...293L..65X,
1996A&A...306...61B, 2000ApJ...528..799W, 2005ApJ...619L..51B}.
Throughout this paper, the IRX is defined in the form of a logarithm:
$\mathrm{IRX} \equiv \log(L(\mathrm{IR})/L(\mathrm{FUV}))$, where
L(IR) is luminosity of infrared thermal radiation of dust and L(FUV)
is far-ultraviolet monochromatic luminosity of starlight (detailed definitions of these terms are presented in Section \ref{measure}). In the
meanwhile, for starburst galaxies, it is believed that the
ultraviolet spectral slope \citep[$\beta$, defined as $f_\lambda
\propto \lambda^{\beta}$, where $1250 \mathrm{Ang} \leq \lambda \leq
2600 \mathrm{Ang}$;][]{1994ApJ...429..582C} has an intrinsic value
\citep[$-2.3 \lesssim \beta \lesssim -2.0$, as suggested
in][]{1994ApJ...429..582C, 1995ApJS...96....9L,
1999ApJ...521...64M}, and any change in this slope results from
wavelength-selective absorption of photons by dust grains. On a
basis of this assumption, the slope of UV spectrum is regarded as
another attenuation estimator \citep{1994ApJ...429..582C,
1997AJ....113..162C}.

The correlation between the two parameters was found by
\citet[][so-called IRX-$\beta$ relation]{1999ApJ...521...64M} with
an investigation based on a local starburst sample, and the relation
was believed to be a sequence indicating dust attenuation. Therefore, the
IRX-$\beta$ relation has great significance for
attenuation correction, since it provides an access to estimate dust
attenuation solely with UV waveband. This prescription has been
widely used at the present time especially for high redshift
galaxies, of which the rest-frame UV observations are available from
ground-based optical telescopes \citep[e.g.,][]{2004ApJ...617..746D,
2007ApJ...670..156D, 2006ApJ...638...72K, 2007ApJS..173..415M,
2009ApJ...705..936B}. Since people usually adopt UV color or UV
luminosity ratio as a surrogate for the traditional UV spectral
slope, hereafter we refer to the relation as the terminology
"IRX-UV" in place of the "IRX-$\beta$".

Observations on extragalactic star-forming
regions and normal galaxies following the discovery of the IRX-UV relation
show considerable dispersion in this relation. The analysis of HII
regions in the Large Magellanic Cloud by
\citet{2002ApJ...577..150B} and the study of normal galaxies by
\citet[][hereafter denoted as K04]{2004MNRAS.349..769K} have
shown that the IRX-UV relation appears to have flatter UV
spectral slopes than starburst galaxies at fixed IRX and covers a wide range; the tight correlation between IRX and $\beta$ for
starburst galaxies has been found in absence, with a considerable
degree of deviation and dispersion from the starburst formula
instead.

Following the launch of the \emph{Galaxy Evolution Explorer}
\citep[\emph{GALEX};][]{2003lgal.conf...10B, 2005ApJ...619L...1M}
and the \emph{Spitzer Space Telescope}
\citep[\emph{Spitzer};][]{2004ApJS..154....1W}, an increasing number
of studies have revealed the deviation in the IRX-UV relation, from
statistics of integrated measurements of galaxies as a whole
\citep[e.g.,][]{2007ApJ...655..863D, 2009ApJ...703..517D,
2007ApJS..173..185G} to spatially resolved studies of individual
galaxies \citep[e.g.,][]{2004ApJS..154..215G, 2005ApJ...633..871C,
2007ApJS..173..572T}. It is worth mentioning that, with the
combination of larger and more comprehensive UV and IR all sky
survey data, \citet{2005ApJ...619L..51B} have found the similar
offset and scatter from the starburst relation in the IRX-UV
diagram, while a few number of IR-bright galaxies appear to have
higher IRXs than starburst galaxies at given UV spectral slopes. The
IRX-rising phenomenon has been confirmed by studies of
(ultra-)luminous infrared galaxies, and adds an additional
question to the IRX-UV issue \citep{2002ApJ...568..651G,
2010ApJ...715..572H}. In order to improve the IRX-UV calibration to
dust attenuation and to provide better insights into the nature of
galactic UV and IR properties, it is necessary to discover the
origin of the deviation in the IRX-UV relation.

A variety of SFH of different stellar populations have been
considered as one main cause of the deviation, and in simple cases
this deviation is indicated by stellar population age. In K04 work, the authors have
compiled a sample of galaxies with a wide coverage of star formation
activities from starburst to quiescent, and employed
the spectral index $D_n(4000)$ to trace different stellar
populations. Via the analysis of perpendicular distances of the
galaxies on the IRX-UV diagram from the empirical best-fitting curve
for starburst galaxies (the starburst empirical curve is
parameterized by Equation (2) in K04 paper), they have confirmed
that variations in stellar populations dominate the dispersion in
the IRX-UV relation. The products of stellar population synthesis
modeling in K04 work have shown that aging of stellar populations
would yield redder unattenuated UV colors (i.e., flatter intrinsic UV
spectral slopes), and the effect of stellar population age on the
IRX-UV relation is depicted as a series of sequences in parallel
with the starburst empirical relation (Figure 4 in K04 paper). Thus,
in addition to dust attenuation, stellar population age has been
considered as the second parameter in the IRX-UV function.
Nevertheless, there has been lack of strong evidence to demonstrate
such an age effect, and in many cases it is ascribed to the fact
that the adopted observational tracers are inappropriate, or the age
effects are weak compared with other potential effects and hence can
be easily masked \citep{2005ApJ...619L..55S, 2006ApJ...637..242C,
2007ApJS..173..392J}. Despite these disappointments, studies of
radial profiles for individual galaxies have disclosed a weak trend
of the age-sensitive color FUV $-$ 3.6 $\mu$m in the IRX-UV
relation, which is in encouraging support of the role stellar
population age plays in the IRX-UV relation
\citep{2009ApJ...701.1965M}. Notwithstanding, the radial profiles
still cannot adequately resolve the mixing of stellar populations
within galaxies.

A recent work on a basis of pixel-by-pixel studies of nearby
galaxies taken by \citet{2012A&A...539A.145B} makes use of the data
obtained from the \emph{Herschel Space Observatory}
\citep{2010A&A...518L...1P} to examine a set of parameters. The
results in their work have illustrated that, intrinsic UV color
plays a predominant role in the IRX-UV relation; the commonly used
age indicators including the spectral index $D_n(4000)$ are in a
poor position to estimate accurate stellar population age and
therefore fail to trace the IRX-UV relation as a function of age; in
addition, the analysis in \citet{2012A&A...539A.145B} has further
predicted a non-negligible contribution from the shape of
attenuation law to the deviation in the IRX-UV relation.
Coincidentally, several numerical simulations have reproduced a
diversity of IRX-UV trends by varying dust-star geometrical
configuration and dust grain properties \citep{2005MNRAS.360.1413B,
2007MNRAS.375..640P, 2008MNRAS.386.1157C}.

In this work, we carry out spatially resolved studies of galaxies
which enable an inspection of galactic sub-structures and allow us
to extract certain sources of interest within each single galaxy
rather than a mixture of different stellar populations
\citep[e.g.,][]{2005ApJ...633..871C, 2006ApJ...648..987P,
2007ApJS..173..572T}, and in this way we can have better separation
between different populations than integrated measurements or radial
profiles of galaxies. This kind of investigation also makes it
possible to compare regions of a similar stellar population in
different galaxies. \citet{2009ApJ...706..553B} have been engaged in
a census of star-forming clusters in eight nearby spiral galaxies.
Further to the approach taken by \citet{2009ApJ...706..553B}, we
focus on all positions in galaxies, not only young clusters, but
also evolved stellar populations in galactic background areas which
are supposed to have different intrinsic UV colors from young
stellar populations. This strategy is a natural way to define
stellar populations and makes it possible to exhibit age signatures
in the IRX-UV diagrams. Up to now, most analyses are based on
integrated measurements of galaxies, while in this work, we take
advantage of the spatially resolved analysis to explore the
determinant of the locations of integrated galaxies in the IRX-UV
diagram. Also, we test the systematic dependence of the IRX-UV
relation on FUV luminosity and radial distance, and present
two-dimensional maps of UV color and IRX to provide a spatial
insight into the IRX-UV relation.

The remainder of this paper is outlined as follows. In Section
\ref{data}, we describe the compilation of the sample, the data
processing and the photometric measurements, and the modeling of
spectral synthesis; in Section \ref{result1}, we present the
resulting IRX-UV diagrams for the extracted regions within the
galaxies in our sample; in Section \ref{result2}, we examine the
systematic dependance of the IRX-UV relations on luminosity and
radial distance; in Section \ref{2-D map}, we provide spatial
distributions of UV color and IRX; Section \ref{disc} is concerned
with relevant discussion and interpretation about the presented
results; finally, we summarize the results and their
implications in Section \ref{sum}.

\section{DATA AND MEASUREMENTS}\label{data}

\subsection{Sample of Galaxies}

The sample of galaxies in this work is compiled from the
\emph{Spitzer} Infrared Nearby Galaxies Survey
\citep[SINGS,][]{2003PASP..115..928K}. The SINGS project has
observed imageries of IR emission components for a sample of 75
nearby galaxies in the wavelength range from 3.6 $\mu$m to 160
$\mu$m. The UV imaging data of the SINGS galaxies have been obtained
as a part of the Ultraviolet Atlas of \emph{GALEX} Nearby Galaxies
Survey \citep{2003lgal.conf...10B, 2007ApJS..173..185G}. Combination
of the both surveys offers an ideal multi-wavelength observational
archive of nearby galaxies, and make it possible for spatially
resolved studies of galaxies.

The \emph{GALEX} observation works at two ultraviolet wavelength
bands: far-ultraviolet (FUV,
$\lambda_\mathrm{eff}=1516\mathrm{Ang}$) and near-ultraviolet (NUV,
$\lambda_\mathrm{eff}=2267\mathrm{Ang}$). The \emph{GALEX} data can
be retrieved on its data release
website.\footnote{\url{http://galex.stsci.edu/}} Detailed
descriptions of the \emph{GALEX} mission and the instruments are
provided in \citet{2005ApJ...619L...1M}. The SINGS data release
includes \emph{Spitzer} imaging at seven bandpasses with two
channels for each of the galaxies: 3.6, 4.5, 5.8, 8.0 $\mu$m with
IRAC, and 24, 70, 160 $\mu$m with MIPS. Readers are referred to
\citet{2004ApJS..154....1W} for a general introduction of the
\emph{Spitzer} telescope, \citet{2004ApJS..154...10F} for the IRAC
instrument, and \citet{2004ApJS..154...25R} for the MIPS instrument.
The SINGS data can be obtained on the \emph{Spitzer} data
distribution
service.\footnote{\url{http://data.spitzer.caltech.edu/popular/sings/}}

\begin{figure}[!ht]
\centering
\vspace*{-10mm}
\includegraphics[width=\columnwidth]{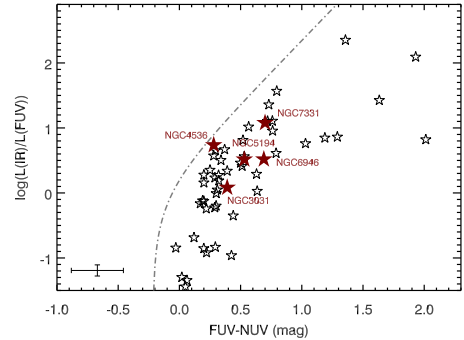}
\vspace*{-55mm}
\caption{IRX vs. $\mathrm{FUV}-\mathrm{NUV}$ for integrated 
measurements of D07 SINGS galaxies. 
The five galaxies studied in this paper are labeled as the brown 
filled stars. The grey dot-dashed line shows K04 empirical curve 
for starburst galaxies. Error bar at the lower left corner
shows the median uncertainty for the sample.} 
\label{IRXUV_D07}
\end{figure}

The approach of this work is to measure all positions within
individual galaxies, and divide various stellar populations
thereinto. For this purpose, large spiral galaxies containing a
sufficient number of well-resolved subregions are required as
targets; on the other hand, in order to investigate distributions of
galaxies in the IRX-UV diagram, we need a sample of galaxies with a
diversity of the IRX-UV locations. Taking the two criteria into
consideration, we select five galaxies from SINGS galaxy sample:
NGC~3031 (M81), NGC~4536, NGC~5194 (M51a), NGC~6946, and NGC~7331.
In this sample, three galaxies have special characters in central
areas: NGC~3031 contains a huge bulge comprising stellar populations
with age older than 8 Gyr \citep{2000AJ....119.2745K}, NGC~4536
performs intensive nuclear star-forming activities
\citep{2005ApJ...630..837J}, and NGC~7331 hosts a ring structure
containing materials of molecular gas and dust
\citep{2004ApJS..154..204R}. In this work, we also aim to
investigate the roles of these special substructures of the IRX-UV
locations of their host galaxies. Figure \ref{IRXUV_D07} displays
the IRX-UV diagram for all the SINGS galaxies with the integrated
measurements quoted from \citet[][hereafter denoted as
D07]{2007ApJ...655..863D}, and the galaxies selected for this work
are labeled. In our sample, NGC~4536 is the closest to the starburst
empirical curve; NGC~6946 is the farthest one; NGC~7331 populates at
a high level in the diagram; NGC~3031 lies near the bottom; and
NGC~5194 ranks at a moderate level. The total IR luminosity is
estimated with 8 $\mu$m and 24 $\mu$m luminosities using the
calibration in \citet[][hereafter denoted as
C05]{2005ApJ...633..871C}. This calibration is employed for
estimation of total IR luminosities throughout this paper, and we
will offer an description in the below section. Table \ref{sample}
presents the basic properties of the galaxies in our sample.

\begin{deluxetable*}{lccccrccc}
\tabletypesize{\scriptsize} \tablecaption{Basic Properties of
Galaxies} \tablewidth{0pc} \tablehead{ \colhead{Name} &
\colhead{R.A.\tablenotemark{a}} & \colhead{Dec.\tablenotemark{a}} &
\colhead{Optical Morphology\tablenotemark{a}} &
\colhead{D25\tablenotemark{a}} &
\colhead{Distance\tablenotemark{b}}&
\colhead{M$_{\mathrm{Opt}}$\tablenotemark{a}} &
\colhead{E(B-V)$_{\mathrm{GAL}}$\tablenotemark{b,c}} & \colhead{SFR\tablenotemark{a}} \\
\colhead{} & \colhead{(J2000.0)} & \colhead{(J2000.0)} &
\colhead{} & \colhead{(arcmin)} & \colhead{(Mpc)} & \colhead{(mag)} & \colhead{(mag)} & \colhead{($M_{\odot}~yr^{-1}$)} \\
} \startdata
NGC~3031~(M81) & 09~55~33.2 & +69~03~55 & SAab & 26.9~$\times$~14.1 & 3.7 & -21.2 & 0.080 & 1.1 \\
NGC~4536 & 12~34~27.0 & +02~11~17 & SABbc & 7.6~$\times$~3.2 & 15.0 & -20.8 & 0.018 & 3.7 \\
NGC~5194~(M51a) & 13~29~52.7 & +47~11~43 & SABbc & 11.2~$\times$~6.9 & 7.7 & -21.4 & 0.035 & 5.4 \\
NGC~6946 & 20~34~52.3 & +60~09~14 & SABcd & 11.5~$\times$~9.8 & 5.6 & -21.3 & 0.342 & 2.2 \\
NGC~7331 & 22~37~04.1 & +34~24~56 & SAb & 10.5~$\times$~3.7 & 14.7 & -21.8 & 0.091 & 4.2 \\
\enddata
\tablenotetext{a}{~Data obtained from \citet{2003PASP..115..928K}.}
\tablenotetext{b}{~Data obtained from the NASA/IPAC Extragalactic
Database.} \tablenotetext{c}{~Data obtained from
\citet{1998ApJ...500..525S}.} \label{sample}
\end{deluxetable*}

\subsection{Image Processing and Aperture Photometry}\label{measure}

Throughout this paper, we derive the total IR luminosity from 8.0
$\mu$m and 24 $\mu$m luminosities by adopting the calibration in
\citet{2005ApJ...633..871C}, where the 8.0 $\mu$m luminosity has
been converted into dust-only emission by subtracting stellar
contribution. The 3.6 $\mu$m images are employed as the reference of
stellar emission by using the scale factor of 0.37 provided in
\citet{2007ApJS..173..572T}. Eventually, we have four-band images
taken with the \emph{GALEX} and \emph{Spitzer} for this study: FUV,
NUV, 8.0 $\mu$m (dust only), and 24 $\mu$m. All these images have
been registered to the same pixel scale and coordinate. The point
spread function of each image has been convolved to match the 24
$\mu$m resolution (FHWM $\sim5.7\arcsec$).

In this work, we design the strategy to measure compact clusters and
background diffuse regions in galaxies respectively, and therefore
attempt to classify young and evolved populations correspondingly.
In order to extract as simplex young populations within clusters as
possible, it is necessary to exclude diffuse emission of local
underlying background at common positions. We mask all the flux
peaks above 3$\sigma$ background level in different images. The
subtraction of local background has been conducted by the
median-filtering approach, that is, for each flux-peak-masked image,
every pixel value is replaced by the median value of all the pixels
enclosed by a square or rectangular window centered on this pixel.
The window size is fixed for every one galaxy but various for
different galaxies. This galaxy-dependent scale is determined by two
criteria: the window size should be large enough to enclose
sufficient background pixels in the flux-peak-masked images, and in
the meanwhile the window cannot extend beyond local areas. In this
situation, we define the final window size as a compromise by manual
adjustment. By means of this approach, we produce a group of
filtered images which represent the local background emission for
each waveband image. This background-fitting procedure is taken with
the MEDIAN program in the IRAF software\footnote{IRAF (the Image
Reduction and Analysis Facility) is a general purpose software
system for the reduction and analysis of astronomical data, and
distributed by the National Optical Astronomy Observatories which is
operated by the Association of Universities for Research in
Astronomy (AURA), Inc. under cooperative agreement with the National
Science Foundation.}. As a result, there are two sets of images for
photometry in this work: the one for which both of global background
and local background have been subtracted is employed to measure
clusters, and the other for which only global background has been
subtracted is applied to photometry for background diffuse regions.

\begin{figure*}[!ht]
\centering
\includegraphics[width=0.9\textwidth]{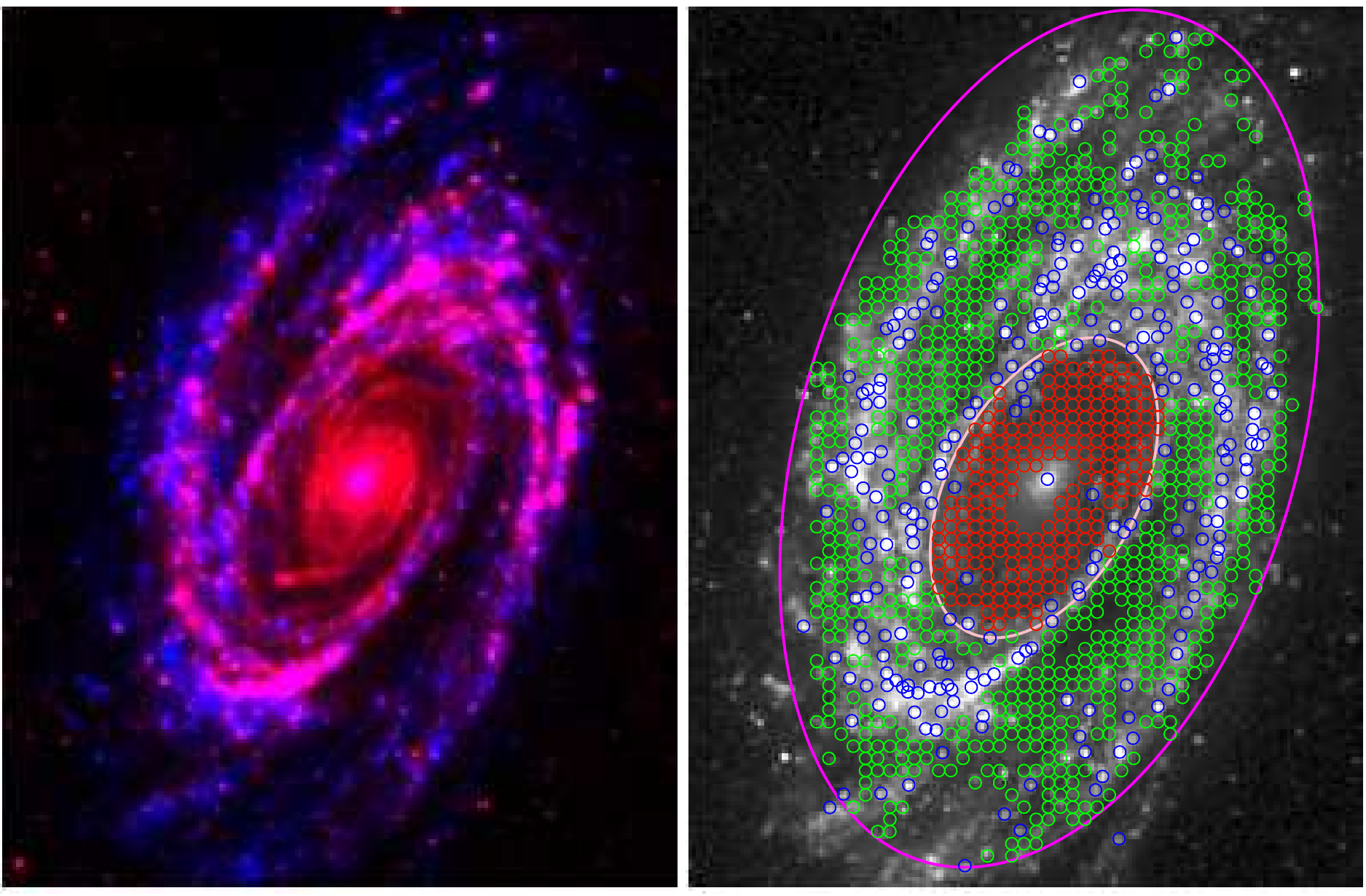}
\caption{Left: Two-color composite
image of NGC~3031. The FUV (blue) and 24 $\mu$m (red)
images are retrieved from \emph{GALEX} and \emph{Spitzer}
observations respectively. Right: \emph{GALEX} FUV image for
NGC~3031. Blue circles enclose UV clusters in the galaxy; green
circles enclose local background regions in disk area and red
circles enclose bulge background regions. The two ellipses show the
apertures use for integrated photometry. The larger magenta ellipse
shows the aperture for the entire galaxy and the pink ellipse is for
the bulge area. The size of the both pictures is $15.6'\times20.4'$.
North is up and east is to the left.} \label{aper_M81}
\end{figure*}  

Aperture photometry is performed in the FUV, NUV, 8 $\mu$m, and 24
$\mu$m images for the galaxies in our sample. We employ circular
apertures in photometry for galactic subregions. The aperture size
for each galaxy is determined mainly by the spatial resolution of
the 24 $\mu$m images: the photometric apertures should be large
enough to adequately enclose the resolved objects, and meanwhile
small enough to avoid as much emission from other adjacent sources
as possible. As a consequence, the radii of the apertures are
galaxy-dependent, but for each one galaxy they are fixed in all
images: 8.5 arcsec for NGC~3031 and NGC~6946, and 6.8 arcsec for
NGC~4536, NGC~5194, and NGC~7331. Correspondingly, the physical
scales of the radii at the distances of the galaxies are 152 pc for
NGC~3031, 495 pc for NGC~4536, 254 pc for NGC~5194, 231 pc for
NGC~6946, and 485 pc for NGC~7331, approximately. We detect emission
peaks by the SExtractor software \citep{1996A&AS..117..393B} in the
\emph{GALEX} FUV images, and these detected UV-emitting sources are
defined as UV clusters representing young stellar populations inside
galaxies. Fluxes of the UV clusters are extracted with the
photometric apertures from the images for which both of global
background and local background have been subtracted. Photometry for
regions in galactic local background which represents evolved
stellar populations is performed in the global-background-subtracted
images. We place photometric apertures for local background regions
on areas between the UV cluster apertures, for instance, on the
bulge, inter spiral arms, and outskirts, to cover as much local
background as possible, and meanwhile to avoid overlap with each
other. The measured UV clusters with fluxes above 3$\sigma$ level of
local background deviation in the
global-plus-local-background-subtracted images are selected in this
work, and likewise, the measured local background regions with
fluxes above 3$\sigma$ level of global background deviation in the
global-background-subtracted images are adopted. Integrated fluxes
of entire galaxies are measured by employing large elliptical
apertures of which major and minor axes are determined by eye to
enclose the main body of each one galaxy. We also extract photometry
of circular or elliptical apertures for galactic center areas.

\begin{figure}
\centering
\includegraphics[width=\columnwidth]{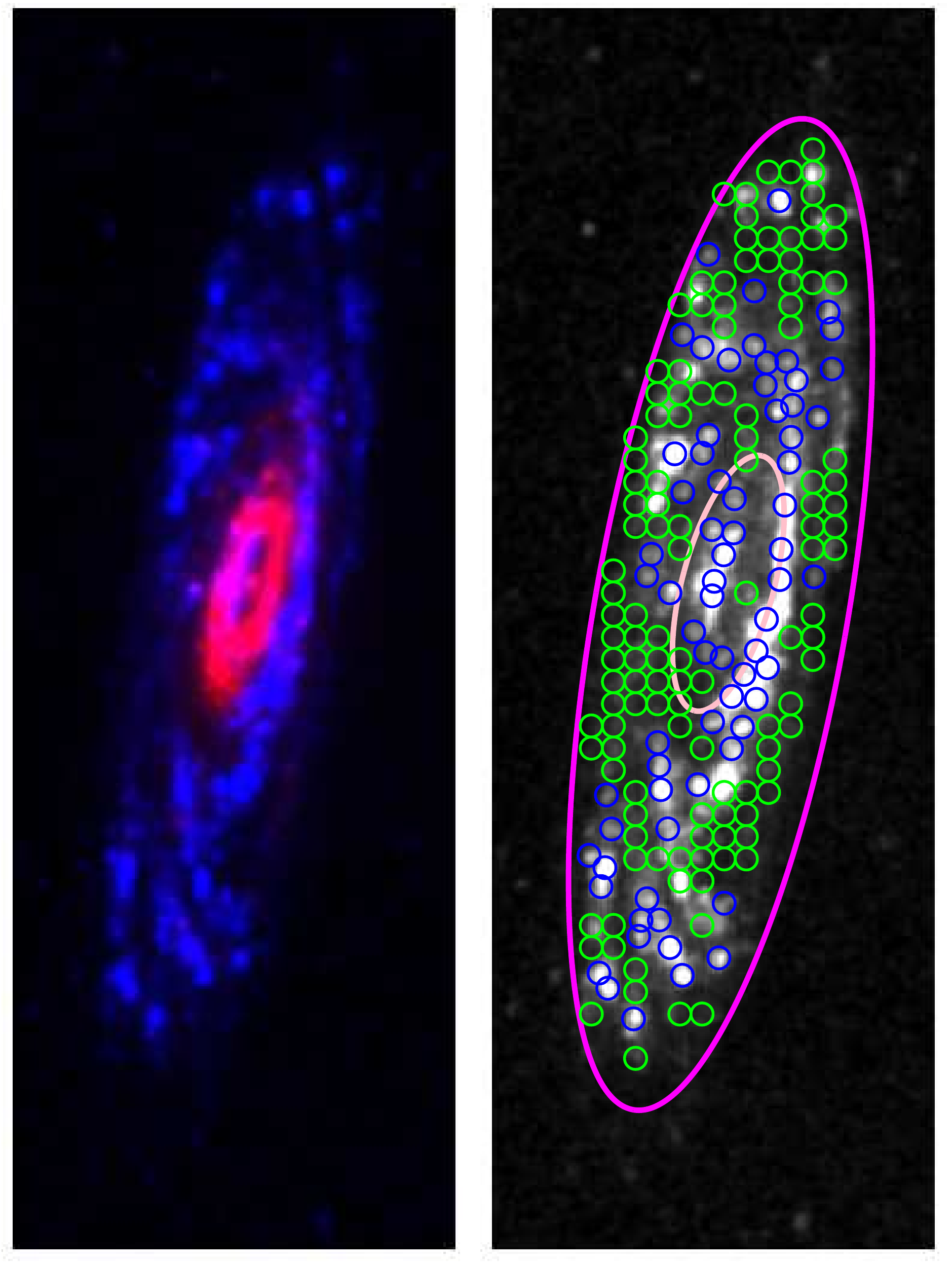}
\caption{Left: Two-color composite
image of NGC~7331. The FUV (blue) and 24 $\mu$m (red)
images are retrieved from \emph{GALEX} and \emph{Spitzer}
observations respectively. Right: \emph{GALEX} FUV image for
NGC~7331. Blue circles enclose UV clusters in the galaxy, and green
circles enclose local background regions in disk area. The two
ellipses show the apertures used for integrated measurement. The
larger magenta ellipse is for the entire galaxy and the pink ellipse
is used for the inner ring area. The size of the both pictures is
$4.5'\times12.7'$. North is up and east is to the left.}
\label{aper_NGC7331}
\end{figure}

Figure \ref{aper_M81} depicts the appearance of the galaxy NGC~3031
from the UV and IR points of view and the photometric apertures
applied to this galaxy as an example. Since for this galaxy the
bulge can be clearly resolved at all wavebands which is believed as
the oldest part of the galaxy, we further divide the sampled local
background regions into disk regions and bulge regions to
distinguish the oldest populations in this galaxy. The left panel of
this figure shows a two-color composite image of NGC~3031, where we
can see FUV and 24 $\mu$m emission radiate predominantly along the
spiral arms, and in the bulge the FUV starlight mostly vanishes but
the 24 $\mu$m dust emission retains a considerable amount. In the
right panel of Figure \ref{aper_M81}, the photometric apertures employed for the UV clusters,
the local background regions in disk and bulge, the whole bulge
area, and the entire galaxy are illustrated. Similarly, we present
the two-color composite image and the exhibition of apertures for
NGC~7331 in Figure \ref{aper_NGC7331}. This galaxy hosts a dust ring
in the center, and we can see this prominent character in the left
panel of Figure \ref{aper_NGC7331}.

The FUV and NUV photometric fluxes are corrected for Galactic
foreground extinction by using the conversion factors provided in
\citet{2007ApJS..173..185G}:
$A_{\mathrm{FUV}}=7.9E(\mathrm{B}-\mathrm{V})_{\mathrm{GAL}}$ and
$A_{\mathrm{NUV}}=8.0E(\mathrm{B}-\mathrm{V})_{\mathrm{GAL}}$, and
the color excess of Galactic extinction
$E(\mathrm{B}-\mathrm{V})_{\mathrm{GAL}}$ of each galaxy in our
sample is obtained from the NASA/IPAC Extragalactic
Database\footnote{\url{http://ned.ipac.caltech.edu/}} where the
\citet{1998ApJ...500..525S} Galactic dust map is quoted, with the
ratio of total to selective extinction $R_\mathrm{V}=3.1$ (note that
$E(\mathrm{B}-\mathrm{V})_{\mathrm{GAL}}$ is independent on
$R_\mathrm{V}$). In this paper, we use $\mathrm{FUV}-\mathrm{NUV}$
color as a surrogate for the UV spectral slope $\beta$, on a basis
of the conversion between UV spectral slope and UV photometric flux
ratio calibrated in K04 paper. Aperture corrections are applied to
the 8 and 24 $\mu$m photometric fluxes following the guidance of the
IRAC and MIPS handbooks provided in the \emph{Spitzer} Science
Center\footnote{\url{http://ssc.spitzer.caltech.edu/}}. The
parameter IRX is calculated according to its definition
$\mathrm{IRX} \equiv \log(L(\mathrm{IR})/L(\mathrm{FUV}))$, where
$L(\mathrm{FUV})={\nu}L_{\nu}(\mathrm{FUV})$, and L(IR) ($\equiv$
L(3$-$1100 $\mu$m) as defined in C05 paper) is derived from 8 $\mu$m
and 24 $\mu$m luminosities via C05 calibration:

\begin{equation}
\log(L(\mathrm{IR}))=
\log(L(24\mu \mathrm{m}))+0.908+0.793\log(L_{\nu}(8\mu \mathrm{m})/L_{\nu}(24\mu \mathrm{m})).
\label{C05_cali}
\end{equation}

Equation (\ref{C05_cali}) is calibrated on a basis of young
star-forming regions \citep[see][]{2005ApJ...633..871C}, and thus
has the potential to underestimate IR luminosity when evolved
stellar populations account for a considerable amount of
dust-heating. In the Appendix, we examine the possible bias in this calibration
and the impact it has on our conclusions.

In this work, uncertainties assigned to the photometric measurements
are estimated as a quadratic sum of the background deviation and
calibration uncertainties in relevant images. The background
deviation is introduced by the background subtraction process. For
the photometry of the UV clusters, the background deviation is
calculated from the global-plus-local-background-subtracted images;
for the local background regions and the integrated galaxies, the
background deviation is derived from the
global-background-subtracted images. The calibration uncertainties
quoted in the quadrature are 0.05 for \emph{GALEX} FUV magnitude and
0.03 for \emph{GALEX} NUV magnitude resulting in an uncertainty of
about 0.06 mag in $\mathrm{FUV}-\mathrm{NUV}$ \citep[$1\sigma$
errors,][]{2007ApJS..173..682M}, 10\% for IRAC 8 $\mu$m flux
\citep{2007ApJ...655..863D}, and 4\% for MIPS 24 $\mu$m flux
\citep{2007PASP..119..994E}.

The measured luminosities from the aperture photometry are listed in
Tables \ref{phot_clu} and \ref{phot_bkg} for the UV clusters and the
local background regions respectively.

\begin{deluxetable*}{lcccccc}
\tabletypesize{\scriptsize} 
\tablecaption{Aperture Photometry of the UV Clusters}
\tablewidth{0pc} \tablehead{ \colhead{ID} & \colhead{R.A.\tablenotemark{a}} & 
\colhead{Dec.\tablenotemark{a}} & \colhead{L(FUV)\tablenotemark{b,c}} &
\colhead{L(NUV)\tablenotemark{b,c}} & \colhead{L(8
$\mu$m-dust)\tablenotemark{b,d}} & \colhead{L(24
$\mu$m)\tablenotemark{b,d}} \\
\colhead{(Cluster Number)} & \colhead{(J2000.0)} & \colhead{(J2000.0)} & \colhead{(ergs~s$^{-1}$)}
& \colhead{(ergs~s$^{-1}$)} & \colhead{(ergs~s$^{-1}$)} &
\colhead{(ergs~s$^{-1}$)} \\
} \startdata
NGC~3031-CLU001 & 148.747 & 69.236 & 4.41e+39~$\pm$~2.19e+38 & 2.98e+39~$\pm$~1.19e+38 & 1.01e+39~$\pm$~1.14e+38 & 8.36e+38~$\pm$~4.44e+37 \\
NGC~3031-CLU002 & 148.853 & 69.219 & 1.11e+40~$\pm$~5.27e+38 & 7.46e+39~$\pm$~2.26e+38 & 4.31e+38~$\pm$~6.83e+37 & 1.64e+38~$\pm$~2.99e+37 \\
NGC~3031-CLU003 & 148.755 & 69.216 & 6.34e+39~$\pm$~3.07e+38 & 3.97e+39~$\pm$~1.40e+38 & 4.29e+39~$\pm$~4.32e+38 & 2.97e+39~$\pm$~1.22e+38 \\
\nodata & & & & & & \\
NGC~4536-CLU001 & 188.576 & 2.223 & 3.03e+40~$\pm$~1.62e+39 & 2.27e+40~$\pm$~9.08e+38 & 1.04e+40~$\pm$~2.31e+39 & 4.28e+39~$\pm$~1.19e+39 \\
NGC~4536-CLU002 & 188.580 & 2.221 & 8.82e+40~$\pm$~4.23e+39 & 6.02e+40~$\pm$~1.81e+39 & 2.40e+40~$\pm$~3.16e+39 & 1.21e+40~$\pm$~1.27e+39 \\
NGC~4536-CLU003 & 188.575 & 2.214 & 4.18e+40~$\pm$~2.12e+39 & 2.95e+40~$\pm$~1.05e+39 & 3.64e+40~$\pm$~4.19e+39 & 2.06e+40~$\pm$~1.44e+39 \\
\nodata & & & & & & \\
NGC~5194-CLU001 & 202.481 & 47.111 & 6.75e+39~$\pm$~1.01e+39 & 3.95e+39~$\pm$~9.41e+38 & 1.25e+40~$\pm$~2.36e+39 & 4.04e+39~$\pm$~1.19e+39 \\
NGC~5194-CLU002 & 202.515 & 47.264 & 3.03e+40~$\pm$~1.72e+39 & 2.29e+40~$\pm$~1.13e+39 & 2.78e+40~$\pm$~3.42e+39 & 1.96e+40~$\pm$~1.41e+39 \\
NGC~5194-CLU003 & 202.517 & 47.260 & 3.46e+40~$\pm$~1.89e+39 & 2.78e+40~$\pm$~1.22e+39 & 4.29e+40~$\pm$~4.73e+39 & 3.50e+40~$\pm$~1.83e+39 \\
\nodata & & & & & & \\
NGC~6946-CLU001 & 308.719 & 60.213 & 1.11e+41~$\pm$~5.29e+39 & 8.98e+40~$\pm$~3.08e+39 & 3.94e+40~$\pm$~4.01e+39 & 2.68e+40~$\pm$~1.14e+39 \\
NGC~6946-CLU002 & 308.727 & 60.213 & 8.61e+40~$\pm$~4.14e+39 & 6.47e+40~$\pm$~2.55e+39 & 1.20e+40~$\pm$~1.44e+39 & 8.23e+39~$\pm$~5.17e+38 \\
NGC~6946-CLU003 & 308.730 & 60.210 & 4.40e+40~$\pm$~2.23e+39 & 3.69e+40~$\pm$~2.06e+39 & 6.76e+39~$\pm$~1.04e+39 & 2.53e+39~$\pm$~4.12e+38 \\
\nodata & & & & & & \\
NGC~7331-CLU001 & 339.257 & 34.481 & 3.55e+40~$\pm$~1.72e+39 & 2.71e+40~$\pm$~9.22e+38 & 2.07e+40~$\pm$~3.43e+39 & 1.00e+40~$\pm$~1.42e+39 \\
NGC~7331-CLU002 & 339.271 & 34.472 & 9.75e+39~$\pm$~5.94e+38 & 7.90e+39~$\pm$~5.66e+38 & 1.43e+40~$\pm$~3.08e+39 & 4.59e+39~$\pm$~1.38e+39 \\
NGC~7331-CLU003 & 339.246 & 34.462 & 1.54e+40~$\pm$~8.16e+38 & 1.16e+40~$\pm$~6.14e+38 & 2.38e+40~$\pm$~3.62e+39 & 4.86e+39~$\pm$~1.38e+39 \\
\nodata & & & & & & \\
\enddata
\tablecomments{The aperture radius is 8.5$\arcsec$ for NGC~3031 and
NGC~6946, and 6.8$\arcsec$ for NGC~4536, NGC~5194, and NGC~7331.
This table is available in its entirety in the online journal. A
portion is shown here for guidance regarding its form and content.}
\tablenotetext{a}{~Position of the apertures on the sky, in units of degree.}
\tablenotetext{b}{~Luminosities measured after local background
subtraction.} \tablenotetext{c}{~Ultraviolet luminosities corrected
for Galactic foreground extinction.} \tablenotetext{d}{~Infrared
luminosities corrected for aperture effects.} \label{phot_clu}
\end{deluxetable*}

\begin{deluxetable*}{ccccccc}
\tabletypesize{\scriptsize} 
\tablecaption{Aperture Photometry of the Local Background Regions}
\tablewidth{0pc} 
\tablehead{ \colhead{ID} & \colhead{R.A.\tablenotemark{a}} & 
\colhead{Dec.\tablenotemark{a}} & \colhead{L(FUV)\tablenotemark{b,c}} &
\colhead{L(NUV)\tablenotemark{b,c}} & \colhead{L(8
$\mu$m-dust)\tablenotemark{b,d}} & \colhead{L(24
$\mu$m)\tablenotemark{b,d}} \\
\colhead{(Region Number)} & \colhead{(J2000.0)} & \colhead{(J2000.0)} & \colhead{(ergs~s$^{-1}$)}
& \colhead{(ergs~s$^{-1}$)} & \colhead{(ergs~s$^{-1}$)} &
\colhead{(ergs~s$^{-1}$)} \\
}
\startdata
NGC~3031-BKG001 & 148.953 & 68.919 & 1.64e+39~$\pm$~7.83e+37 & 1.16e+39~$\pm$~1.10e+38 & 3.21e+38~$\pm$~5.36e+37 & 5.66e+37~$\pm$~1.11e+37 \\
NGC~3031-BKG002 & 148.926 & 68.919 & 9.12e+38~$\pm$~4.51e+37 & 7.09e+38~$\pm$~1.07e+38 & 4.32e+38~$\pm$~6.09e+37 & 1.35e+38~$\pm$~1.21e+37 \\
NGC~3031-BKG003 & 148.926 & 68.924 & 7.42e+38~$\pm$~3.75e+37 & 6.10e+38~$\pm$~1.07e+38 & 3.28e+38~$\pm$~5.40e+37 & 1.26e+38~$\pm$~1.20e+37 \\
\nodata & & & & & & \\
NGC~4536-BKG001 & 188.653 & 2.144 & 1.92e+40~$\pm$~9.14e+38 & 1.29e+40~$\pm$~7.11e+38 & 4.42e+39~$\pm$~4.90e+38 & 1.38e+39~$\pm$~1.58e+38 \\
NGC~4536-BKG002 & 188.672 & 2.148 & 5.30e+39~$\pm$~2.77e+38 & 5.96e+39~$\pm$~6.35e+38 & 5.30e+39~$\pm$~5.71e+38 & 3.82e+39~$\pm$~2.13e+38 \\
NGC~4536-BKG003 & 188.653 & 2.148 & 2.69e+40~$\pm$~1.27e+39 & 1.93e+40~$\pm$~8.17e+38 & 4.55e+39~$\pm$~5.03e+38 & 7.35e+38~$\pm$~1.51e+38 \\
\nodata & & & & & & \\
NGC~5194-BKG001 & 202.475 & 47.113 & 3.22e+39~$\pm$~1.58e+38 & 2.70e+39~$\pm$~1.02e+38 & 5.87e+39~$\pm$~5.94e+38 & 2.21e+39~$\pm$~9.72e+37 \\
NGC~5194-BKG002 & 202.469 & 47.113 & 4.40e+39~$\pm$~2.12e+38 & 3.08e+39~$\pm$~1.10e+38 & 8.38e+39~$\pm$~8.43e+38 & 3.45e+39~$\pm$~1.44e+38 \\
NGC~5194-BKG003 & 202.463 & 47.113 & 2.52e+39~$\pm$~1.27e+38 & 1.97e+39~$\pm$~8.74e+37 & 4.82e+39~$\pm$~4.91e+38 & 2.11e+39~$\pm$~9.37e+37 \\
\nodata & & & & & & \\
NGC~6946-BKG001 & 308.680 & 60.078 & 1.23e+40~$\pm$~7.71e+38 & 1.29e+40~$\pm$~3.08e+39 & 3.65e+39~$\pm$~5.36e+38 & 1.72e+39~$\pm$~9.66e+37 \\
NGC~6946-BKG002 & 308.765 & 60.083 & 2.09e+40~$\pm$~1.11e+39 & 1.72e+40~$\pm$~3.10e+39 & 2.54e+39~$\pm$~4.68e+38 & 1.02e+39~$\pm$~7.92e+37 \\
NGC~6946-BKG003 & 308.680 & 60.083 & 1.14e+40~$\pm$~7.38e+38 & 1.12e+40~$\pm$~3.08e+39 & 9.55e+39~$\pm$~1.03e+39 & 3.65e+39~$\pm$~1.61e+38 \\
\nodata & & & & & & \\
NGC~7331-BKG001 & 339.286 & 34.335 & 7.08e+39~$\pm$~3.56e+38 & 7.38e+39~$\pm$~1.78e+39 & 1.28e+40~$\pm$~1.30e+39 & 3.63e+39~$\pm$~2.04e+38 \\
NGC~7331-BKG002 & 339.295 & 34.342 & 8.20e+39~$\pm$~4.06e+38 & 8.61e+39~$\pm$~1.78e+39 & 9.28e+39~$\pm$~9.56e+38 & 3.40e+39~$\pm$~1.97e+38 \\
NGC~7331-BKG003 & 339.277 & 34.342 & 6.00e+39~$\pm$~3.09e+38 & 6.39e+39~$\pm$~1.77e+39 & 1.76e+40~$\pm$~1.78e+39 & 5.95e+39~$\pm$~2.77e+38 \\
\nodata & & & & & & \\
\enddata
\tablecomments{The aperture radius is 8.5$\arcsec$ for NGC~3031 and
NGC~6946, and 6.8$\arcsec$ for NGC~4536, NGC~5194, and NGC~7331.
This table is available in its entirety in the online journal. A
portion is shown here for guidance regarding its form and content.}
\tablenotetext{a}{~Position of the apertures on the sky, in units of degree.}
\tablenotetext{b}{~Luminosities measured after global background
subtraction.} \tablenotetext{c}{~Ultraviolet luminosities corrected
for Galactic foreground extinction.} \tablenotetext{d}{~Infrared
luminosities corrected for aperture effects.} \label{phot_bkg}
\end{deluxetable*}

\subsection{Stellar Population Synthesis Modeling}

We construct stellar population spectra with a series of ages via
the STARBURST99 library of evolutionary population synthesis
\citep{1999ApJS..123....3L, 2005ApJ...621..695V}. The simulation is
on assumptions of simple stellar populations born with an
instantaneous burst, the solar metallicity of $Z=0.02$, and the
\citet{2002Sci...295...82K} initial mass function (IMF) with
exponents of 1.3 over 0.1$-$0.5 $M_{\odot}$ and 2.3 over 0.5$-$100
$M_{\odot}$ for stellar populations in the modeled spectra. The
assumption of an instantaneous burst is a simplified account for
stellar populations contained in galactic subregions, and this
scenario of simple stellar populations highlights the age signature
in the IRX-UV function. However, in view of any possible variation
in SFH within the measured regions inside galaxies, descriptions
with such a scenario are not sufficient to make realistic estimates
of age. We will discuss more complex SFHs by adopting scenarios of
composite stellar populations in Section \ref{disc}.

All the modeled stellar population spectra are charged with a series
of dust attenuation, and in this procedure we employ the starburst
attenuation law \citep[][hereafter denoted as
C00]{2000ApJ...533..682C} to produce such attenuation with
$R_\mathrm{V}=4.05$. Dust attenuation and stellar population age are
the variables in the products of modeling. C00 starburst attenuation
law is based on the same sample as fitted with K04 empirical
relation, and the reproduced IRX-UV curves are an extension of K04
empirical curve in the age space, i.e., the IRX-UV relation with age
as the second parameter. In this way, we establish an IRX-UV
function with two parameters: dust attenuation and stellar
population age, for the purpose of comparison with observational
data. It is possible for attenuation law to vary with different
environments, and the application of a common attenuation law to
different galaxies and different galactic subregions aims at
simplicity. In the second paper of this series (Paper II, Ye-Wei Mao et al. 2012, in
preparation), we will focus on the issue of the IRX-UV properties
related to attenuation/extinction law.

By convolving the modeled spectra with transmission curves of the
\emph{GALEX} FUV and NUV filters, we obtain FUV and NUV
luminosities, and $\mathrm{FUV}-\mathrm{NUV}$ colors; the total IR
luminosity in the modeling is calculated as a sum of all the
attenuated stellar emission, on a basis of the energy balance
assumption that the attenuated starlight is re-emitted by
interstellar dust as IR continuum of a equal amount. As a result we
reproduce the IRX-UV relation as a function of stellar population
age.

\section{THE IRX-UV RELATION}\label{result1}

\begin{figure}[!ht]
\centering
\vspace*{-10mm}
\includegraphics[width=\columnwidth]{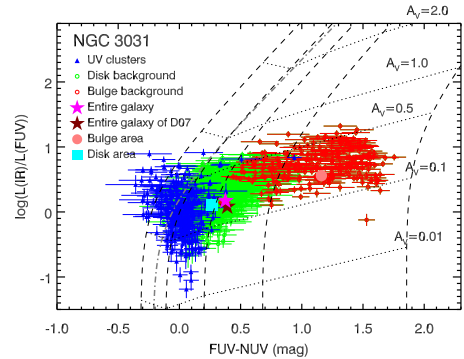}
\vspace*{-55mm}
\caption{IRX vs. $\mathrm{FUV}-\mathrm{NUV}$ for NGC~3031 with the
relational curves reproduced by C00 starburst attenuation law
superimposed. Blue filled triangles symbolize the UV clusters (the
blue circles in Figure \ref{aper_M81}), green open circles represent
the local background regions in disk area (the green circles in
Figure \ref{aper_M81}), and red open circles depict the local
background regions regions in bulge area (the red circles in Figure
\ref{aper_M81}). The integrated measurement of the entire galaxy is
shown as the magenta filled star (the magenta ellipse in Figure
\ref{aper_M81}), the pink filled circle shows the whole bulge area
(the pink ellipse in Figure \ref{aper_M81}), the cyan filled square
symbolizes the disk part of the galaxy, and the brown filled star
presents the integrated galaxy with fluxes cited from the D07 paper.
Black dashed lines describe the model curves sampled with five ages:
2, 8, 100, 300, and 500 Myr, from left to right on the horizontal
axis. Dotted lines connect the points of five constant amounts of
dust attenuation ($A_\mathrm{V}$ = 0.01, 0.1, 0.5, 1.0, and 2.0) on
the model curves of different ages. Grey dot-dashed line is K04
starburst curve. Error bars showing the photometric uncertainties
are plotted as well.} \label{IRXUV_UV_M81}
\end{figure}

With the measurements described above, in this section, we analyze
the IRX-UV diagrams for the galaxies in our sample: NGC~3031,
NGC~4536, NGC~5194, NGC~6946, and NGC~7331. Figures
\ref{IRXUV_UV_M81}$-$\ref{IRXUV_UV_NGC7331} show the IRX-UV diagrams
for the UV clusters and the local background regions inside these
galaxies, with relevant empirical and modeled curves superimposed.
Taking a general view of each of these diagrams we can see that the
data points in each plot compose a continuous locus and spread in a
redder $\mathrm{FUV}-\mathrm{NUV}$ color range or a lower IRX level
than K04 starburst empirical reference. In the following
subsections, we present respective descriptions for these galaxies.

\subsection{NGC~3031 (M81)}\label{3031}

For NGC~3031, we divide the measured subregions into three
populations: UV clusters, disk background regions, and bulge
background regions, representing young, evolved, and the oldest
populations in this galaxy, respectively, as described in the above
section. Figure \ref{IRXUV_UV_M81} shows the IRX-UV distributions
for the three populations, the integrated galaxy, and the integrated
bulge, with K04 empirical curve and the C00-attenuation-modeled
curves for constant ages superimposed. The locus composed of all the
data points presents a large departure from K04 empirical curve, and
appears to have flatter IRXs with increasing
$\mathrm{FUV}-\mathrm{NUV}$ colors. The most remarkable feature in
this figure is the clear separation of different populations. Most
of the data points populate in the IRX range between 0.0 and 1.0,
but the different populations appear to lie in possession of their
own UV color ranges: the UV clusters occupy the
$\mathrm{FUV}-\mathrm{NUV}$ range between -0.3 and 0.2 mag, the disk
background regions spread in the $\mathrm{FUV}-\mathrm{NUV}$ range
of 0.2$-$0.7 mag, and the bulge background regions extend from 0.7
to 1.8 mag in $\mathrm{FUV}-\mathrm{NUV}$, and are thus embedded
in the reddest zone of the panel. The population classification
enables us to obviously see the increasing perpendicular distances
from the starburst empirical curve with aging of stellar
populations, and this behavior is in rough similarity to K04
prediction. The separation of different populations provides a
robust interpretation of the wide spread of the data points with the
effects of stellar population age.

In Figure \ref{IRXUV_UV_M81}, we overplot the IRX-UV curves at five
certain stellar population ages: 2 Myr, 8 Myr, 100 Myr, 300 Myr, and
500 Myr, reproduced by the two-parameter modeling. In the diagram,
the modeled curves are displayed in parallel with K04 empirical
line, and the 8 Myr-age curve overlaps well with this empirical
line. With stellar populations evolving, the relevant IRX-UV curves
shift towards redder $\mathrm{FUV}-\mathrm{NUV}$ colors, due to the
inherently red UV colors of evolved stellar populations. We also
connect constant dust attenuation on each age-fixed curve, and the
tracks appear to be slightly steeper than horizontal (see the dotted
lines in Figure \ref{IRXUV_UV_M81}), which shows that IRX is not
perfectly independent on age. The curves at fixed age and the tracks
of constant dust attenuation are in agreement with the model
products from the same attenuation law in
\citet{2005ApJ...633..871C} and \citet{2009ApJ...706..553B}. The
modeled grid illustrates clearly the behaviors of dust attenuation
and age parameters in the IRX-UV function. The model here is based
on the SFH of an instantaneous burst, and Section \ref{disc} will
present the cases of more complex SFHs.

\begin{deluxetable*}{lcccccc}[!ht]
\tabletypesize{\scriptsize} 
\tablecaption{The Luminosity Contributions of the UV Clusters and
Galaxy Centers to the Integrated Measurements of Galaxies}
\tablehead{ \colhead{} & \colhead{Galaxy} &
\colhead{L(FUV)} & \colhead{L(NUV)} & \colhead{L(8 $\mu$m-dust)} &
\colhead{L(24
$\mu$m)} & \colhead{L(IR)} \\
\colhead{} & \colhead{} & \colhead{($10^{42}$ ergs~s$^{-1})$} &
\colhead{($10^{42}$ ergs~s$^{-1}$)} & \colhead{($10^{42}$ ergs~s$^{-1}$)} &
\colhead{($10^{42}$ ergs~s$^{-1}$)} &
\colhead{($10^{42}$ ergs~s$^{-1}$)} \\
} \startdata
UV~Clusters/Entire~Galaxy & NGC~3031 & 0.272~$\pm$~0.013 & 0.203~$\pm$~0.006 & 0.226~$\pm$~0.023 & 0.351~$\pm$~0.015 & 0.236~$\pm$~0.022 \\
\nodata & NGC~4536 & 0.227~$\pm$~0.011 & 0.203~$\pm$~0.006 & 0.557~$\pm$~0.070 & 0.771~$\pm$~0.040 & 0.592~$\pm$~0.070 \\
\nodata & NGC~5194 & 0.291~$\pm$~0.014 & 0.287~$\pm$~0.008 & 0.329~$\pm$~0.033 & 0.437~$\pm$~0.018 & 0.346~$\pm$~0.033 \\
\nodata & NGC~6946 & 0.242~$\pm$~0.012 & 0.199~$\pm$~0.006 & 0.207~$\pm$~0.021 & 0.254~$\pm$~0.010 & 0.213~$\pm$~0.020 \\
\nodata & NGC~7331 & 0.239~$\pm$~0.011 & 0.217~$\pm$~0.006 & 0.240~$\pm$~0.024 & 0.295~$\pm$~0.012 & 0.249~$\pm$~0.024 \\
\\
Galaxy~Center/Entire~Galaxy & NGC~3031 & 0.087~$\pm$~0.006 & 0.178~$\pm$~0.007 & 0.185~$\pm$~0.026 & 0.352~$\pm$~0.020 & 0.211~$\pm$~0.028 \\
\nodata & NGC~4536 & 0.234~$\pm$~0.016 & 0.270~$\pm$~0.011 & 0.774~$\pm$~0.109 & 0.881~$\pm$~0.050 & 0.795~$\pm$~0.106 \\
\nodata & NGC~5194 & 0.103~$\pm$~0.007 & 0.152~$\pm$~0.006 & 0.216~$\pm$~0.031 & 0.255~$\pm$~0.014 & 0.224~$\pm$~0.030 \\
\nodata & NGC~6946 & 0.021~$\pm$~0.001 & 0.026~$\pm$~0.001 & 0.203~$\pm$~0.029 & 0.313~$\pm$~0.018 & 0.222~$\pm$~0.030 \\
\nodata & NGC~7331 & 0.157~$\pm$~0.010 & 0.265~$\pm$~0.011 & 0.584~$\pm$~0.083 & 0.629~$\pm$~0.036 & 0.593~$\pm$~0.079 \\
\\
Galaxy~Luminosity~(OURs) 
        & NGC~3031 & 4.76~$\pm$~0.224 & 4.50~$\pm$~0.126 & 2.57~$\pm$~0.257 & 0.94~$\pm$~0.038 & 7.08~$\pm$~0.668 \\
\nodata & NGC~4536 & 8.47~$\pm$~0.399 & 7.68~$\pm$~0.216 & 15.0~$\pm$~1.500 & 11.1~$\pm$~0.442 & 47.7~$\pm$~4.490 \\
\nodata & NGC~5194 & 20.2~$\pm$~0.952 & 21.2~$\pm$~0.593 & 27.0~$\pm$~2.700 & 11.6~$\pm$~0.463 & 76.8~$\pm$~7.240 \\
\nodata & NGC~6946 & 18.9~$\pm$~0.890 & 21.9~$\pm$~0.628 & 16.9~$\pm$~1.690 & 8.61~$\pm$~0.344 & 49.8~$\pm$~4.690 \\
\nodata & NGC~7331 & 6.61~$\pm$~0.311 & 8.84~$\pm$~0.251 & 31.8~$\pm$~3.180 & 12.1~$\pm$~0.482 & 88.1~$\pm$~8.310 \\
\\
Galaxy~Luminosity~(D07) 
        & NGC~3031 & 5.23~$\pm$~0.726 & 5.05~$\pm$~0.697 & 2.24~$\pm$~0.637 & 0.95~$\pm$~0.037 & 6.34~$\pm$~1.470 \\
\nodata & NGC~4536 & 8.63~$\pm$~1.200 & 7.52~$\pm$~1.040 & 14.7~$\pm$~2.050 & 11.2~$\pm$~0.454 & 47.1~$\pm$~5.750 \\
\nodata & NGC~5194 & 23.9~$\pm$~3.310 & 26.2~$\pm$~3.630 & 27.5~$\pm$~3.780 & 12.0~$\pm$~0.504 & 78.6~$\pm$~9.530 \\
\nodata & NGC~6946 & 15.2~$\pm$~2.120 & 19.4~$\pm$~2.700 & 17.0~$\pm$~2.330 & 8.93~$\pm$~0.355 & 50.3~$\pm$~6.040 \\
\nodata & NGC~7331 & 7.48~$\pm$~1.040 & 9.59~$\pm$~1.330 & 31.7~$\pm$~4.640 & 13.3~$\pm$~0.764 & 89.6~$\pm$~12.30 \\

\enddata
\label{contri}
\end{deluxetable*}

By comparing the observational data and the model-based
curves in Figure \ref{IRXUV_UV_M81}, we can see that a vast number
of the data points are enclosed in the attenuation range between
$A_V=0.1$ and $A_V=0.5$. Although the separate distributions of
different populations offer an obvious evidence of the age effects
on the IRX-UV relation, the population separation in the diagram
cannot quantitatively correspond to the model indicators. The data
locus of any single population is not in agreement with the modeled
IRX-UV curves with fixed ages. According to the modeled grid, about
a half of the disk background regions have to be partitioned into
the age range younger than 100 Myr and therefore cannot be
age-distinguished from most of the UV clusters in this situation,
while a few number of UV clusters with blue color lie out of the
model coverage. Error bars of the data points are displayed in this
diagram. The typical uncertainties in $\mathrm{FUV}-\mathrm{NUV}$
are $\pm$0.07 mag for the UV clusters and $\pm$0.14 mag for the
local background regions, and in IRX the typical uncertainty is
$\pm$0.06 for either the UV clusters or the local background
regions. We inspect the blue clusters located out of the model
coverage and find the median uncertainty in
$\mathrm{FUV}-\mathrm{NUV}$ is $\pm$0.13 mag and the maximum is
$\pm$0.25 mag. Although these clusters have relatively large errors
which would cause dispersion in the data distribution, the photometric
uncertainties are inadequate to account for the systematic
discrepancy between the data distribution and the model expectation.
Conclusively, the IRX-UV diagram for NGC~3031 shows the impacts of
stellar population age, but there are insufficiencies in the
interpretation of the data distribution even though stellar
population age is addressed as the second parameter in the IRX-UV
function. It should be note that, in this section we characterize
the observational results with the scenario assuming simple stellar
populations with an instantaneous burst, which is an simplification
of stellar populations for the subregions inside galaxies. More
complicated SFHs with exponentially decreasing SFRs will be
discussed in Section \ref{disc}, and offer complementary scenarios
for interpreting the observational properties.

The integrated galaxy of NGC~3031 is located at
$\mathrm{FUV}-\mathrm{NUV}$ = 0.38 $\pm$ 0.06 mag and IRX = 0.17
$\pm$ 0.05, encompassed with the disk background regions, while the
integrated bulge populates at $\mathrm{FUV}-\mathrm{NUV}$ = 1.16
$\pm$ 0.06 mag and IRX = 0.56 $\pm$ 0.05, in large disparity from
the entire galaxy. In Table \ref{contri}, we provide the luminosity
contributions of the UV clusters and the center areas to the
integrated measurements of entire galaxies. From this table, we can
see that the total luminosity of the UV clusters account for 35\% of
24 $\mu$m and less than 30\% of any other waveband integrated
luminosity of NGC~3031. This percentage presentation indicates that
the primary fraction of the integrated luminosities of NGC~3031
comes from the local background including the local background
regions we have measured, the areas between the photometric
apertures we are unable to extract, and the local background at the
positions of the UV clusters we have subtracted. The contributions
of the UV clusters are of less but non-negligible significance,
while the contributions of the whole bulge area are quite trivial,
which further suggests the dominant role of the disk background in
the integrated measurements, and offers an explanation of the
location of the integrated galaxy in the IRX-UV diagram.

\subsection{NGC~4536}\label{4536}

\begin{figure}
\centering
\vspace*{-10mm}
\includegraphics[width=\columnwidth]{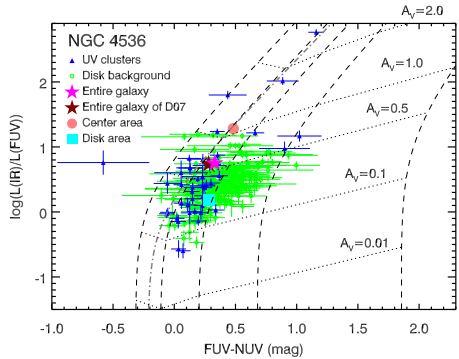}
\vspace*{-55mm}
\caption{IRX vs. $\mathrm{FUV}-\mathrm{NUV}$ of NGC~4536 for the UV
clusters (blue filled triangles), the local background
regions (green open circles), the entire galaxy
(magenta filled star), the center area (pink filled
circle), the disk area (cyan filled square), and the entire
galaxy with fluxes cited from the D07 paper (brown filled
star), with the same curves as shown in Figure \ref{IRXUV_UV_M81}
superimposed. Error bars showing the photometric uncertainties are
plotted as well.} \label{IRXUV_UV_NGC4536}
\end{figure}

Figure \ref{IRXUV_UV_NGC4536} shows the IRX-UV relation for the UV
clusters and the local background regions within the galaxy
NGC~4536. We can see the locus of the UV clusters coincidentally
follows K04 curve with an offset of $\sim$0.3 mag on average towards
redder $\mathrm{FUV}-\mathrm{NUV}$ colors at fixed IRX, and the data
points distribute along the modeled tracks of constant ages. The
studies of this galaxy on the radial profile
\citep{2009ApJ...701.1965M} and on the pixel-by-pixel basis
\citep{2012A&A...539A.145B} have obtained the consistent result: the
distribution of data points for this galaxy is coherent with the
starburst empirical relation. The UV clusters span a wide range in
this diagram approximately from $-$0.1 to 1.2 mag in
$\mathrm{FUV}-\mathrm{NUV}$ and from $-$0.7 to 2.8 in IRX, and the
most number of the UV clusters are enclosed in the age range $<$ 100
Myr and extend over a large scale of the attenuation amount. The
cluster with IRX $\simeq$ 2.8 in this figure is extracted at the
core of this galaxy, where intense star-forming activities occur and
manufacture the strong IR emission by dust-heating. The local
background regions systematically shift to redder
$\mathrm{FUV}-\mathrm{NUV}$ colors than the UV clusters by a mean
factor of $\sim$0.3 mag at fixed IRX. Consequently they extend to even
older ages of over 300 Myr. The offset between the UV clusters and
the local background regions in the diagram presents an obvious
behavior of the age parameter in the modeled IRX-UV function.

The typical uncertainties in $\mathrm{FUV}-\mathrm{NUV}$ are
$\pm$0.07 mag for the UV clusters and $\pm$0.12 for the local
background regions, and in IRX are $\pm$0.08 for the UV clusters and
$\pm$0.06 for the local background regions. There is no effective
influence on the results in consideration of the photometric
uncertainties.

The integrated galaxy lies at $\mathrm{FUV}-\mathrm{NUV}$ = 0.33
$\pm$ 0.06 mag and IRX = 0.75 $\pm$ 0.05, and the galactic center
populates at $\mathrm{FUV}-\mathrm{NUV}$ = 0.48 $\pm$ 0.06 mag and
IRX = 1.28 $\pm$ 0.05. Both of the points abut on the starburst
empirical line as shown in Figure \ref{IRXUV_UV_NGC4536}. Table
\ref{contri} shows that the UV clusters contribute $\sim$20\% of UV
emission and more than a half of IR luminosity of the galaxy. This
large IR contribution is due to the UV clusters in the center
star-forming area. We can see that the whole center area accounts
for even 80\% of the total IR luminosity and 88\% of the 24 $\mu$m
monochromatic luminosity of the galaxy, but the integrated FUV and
NUV luminosities of NGC~4536 are dominated by the disk area. The
contribution of the star-forming center to the integrated galaxy
introduces the adjacency of the integrated galaxy to the starburst
empirical line. As a result the star-forming center plays a primary
role in the integrated IRX-UV feature for NGC~4536.

\subsection{NGC~5194 (M51a)}\label{5194}

\begin{figure}
\centering
\vspace*{-10mm}
\includegraphics[width=\columnwidth]{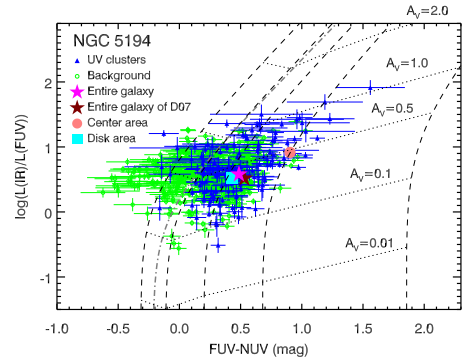}
\vspace*{-55mm}
\caption{IRX vs. $\mathrm{FUV}-\mathrm{NUV}$ of NGC~5194 for the UV
clusters (blue filled triangles), the local background
regions (green open circles), the entire galaxy
(magenta filled star), the center area (pink filled
circle), the disk area (cyan filled square), and the entire
galaxy with fluxes cited from the D07 paper (brown filled
star), with the same curves as shown in Figure \ref{IRXUV_UV_M81}
superimposed. Error bars showing the photometric uncertainties are
plotted as well.} \label{IRXUV_UV_M51}
\end{figure}

In Figure \ref{IRXUV_UV_M51}, the IRX-UV locus for the UV clusters
of NGC~5194 appears to have redder UV colors with increasing IRX and
in consequence a shallower trend than the empirical relation and the
modeled curves. In the two-parameter scenario of model, the UV
clusters with red $\mathrm{FUV}-\mathrm{NUV}$ color driving the
locus shallower fail to reach high attenuation levels, and therefore
leak into the old regime. The typical uncertainties for the UV
clusters are $\pm$0.08 mag in $\mathrm{FUV}-\mathrm{NUV}$ and
$\pm$0.07 in IRX. The uncertainties for the red clusters with
$\mathrm{FUV}-\mathrm{NUV}>1.0$ mag and IRX $>$ 1.0 are relatively
larger: $\sim \pm$0.28 mag in $\mathrm{FUV}-\mathrm{NUV}$ and $\sim
\pm$0.11 in IRX. However, including the photometric uncertainties,
the model description still tends to give an overestimate of age for
the UV clusters, in particular, about a half number of the UV
clusters have to be unreasonably enclosed in the age range of
100$-$300 Myr. This phenomenon shows a discrepancy in characterizing
the IRX-UV relation for NGC~5194 by the two-parameter scenario.

In contrast to NGC~3031 and NGC~4536, the
NGC~5194 background regions are mixed with the UV clusters in the
IRX-UV diagram, and even about a half of the local background
regions show bluer UV colors than the bulk of the UV clusters.
Hence, any age effect disappears. Through a visual inspection in
images, we find that the typical background of diffuse emission is
extremely weak within NGC~5194. Instead, in the areas of inter-arms
and outskirts (the background regions with bluer
$\mathrm{FUV}-\mathrm{NUV}$ color than the UV clusters populate
mainly in outskirts of the galactic body), there are a great number
of UV knots in small size with low luminosity. It is likely for the
local background regions inside this galaxy to experience different
SFHs, where long-term or even continuous star formation is supposed
to take place. In this situation, the measured regions in background
ought to contain a large number of young populations, and as a result,
they should present $\mathrm{FUV}-\mathrm{NUV}$ properties similar to those of the
UV clusters. More complex SFHs will be addressed in our discussion
in Section \ref{disc}.

The quantification of luminosity contributions in Table \ref{contri}
suggests a very dim center area of NG5194 at UV and IR wavelength
bands which accounts for 10\% of FUV, 15\% of NUV, and 22\% of total
IR luminosities of the entire galaxy. The integrated photometry of
this galaxy is dominated by the disk luminosity, and the UV clusters
contribute about 30\% of FUV, NUV, and 8 $\mu$m, and 44\% of 24
$\mu$m, and 35\% of total IR luminosities to the integrated
measurements of NGC~5194, which implies the local background
contributes more than half of the integrated luminosities of the
galaxy at all observational bands. However, in Figure
\ref{IRXUV_UV_M51}, the point of the integrated galaxy is located at
the position of $\mathrm{FUV}-\mathrm{NUV}$ = 0.49 $\pm$ 0.06 mag
and IRX = 0.58 $\pm$ 0.05, not at the average position of the local
background regions. The possible reason is that a number of the
green points in the plot are not good proxies for representing the local
background. They populate in the outskirts of the galactic body, and
thus in a special instance of the background. The background regions
which populate in between cluster-apertures, and at the positions of
the UV clusters are unable to be extracted. These missed regions are
believed to better signify the background properties in integrated
measurements. Provisionally, in Figure \ref{IRXUV_UV_M51} the
extracted inter-arms regions are mixed with the UV clusters and
considered to be relatively more representative.

\begin{figure}
\centering
\vspace*{-10mm}
\includegraphics[width=\columnwidth]{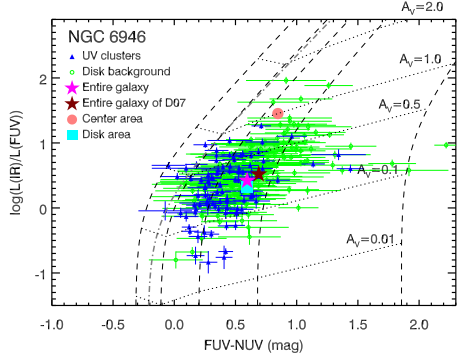}
\vspace*{-55mm}
\caption{IRX vs. $\mathrm{FUV}-\mathrm{NUV}$ of NGC~6946 for the UV
clusters (blue filled triangles), the local background
regions (green open circles), the entire galaxy
(magenta filled star), the center area (pink filled
circle), the disk area (cyan filled square), and the entire
galaxy with fluxes cited from the D07 paper (brown filled
star), with the same curves as shown in Figure \ref{IRXUV_UV_M81}
superimposed. Error bars showing the photometric uncertainties are
plotted as well.} \label{IRXUV_UV_NGC6946}
\vspace*{2mm}
\end{figure}

\subsection{NGC~6946}\label{6946}

The IRX-UV diagram for NGC~6946 is displayed in Figure
\ref{IRXUV_UV_NGC6946}. The local background regions present a shift
towards redder UV color than the UV clusters, even up to
$\mathrm{FUV}-\mathrm{NUV}\sim2.0$ mag. However, the large scatter
in the data distribution ($\sim$0.5 mag in
$\mathrm{FUV}-\mathrm{NUV}$ at fixed IRX) appears of more
significance, and the signature of the age parameter tends to be
faded by the dispersion. The photometric uncertainties for this
galaxy are relatively larger than others in our sample. The typical
$\mathrm{FUV}-\mathrm{NUV}$ uncertainty is $\pm$0.09 mag for the UV
clusters and $\pm$0.25 mag for the local background regions, and the
typical IRX uncertainty is $\pm$0.05 for either the UV clusters or
the local background regions. The photometric uncertainties are one
of the main sources of the scatter. It is worth noting that, the
foreground Galactic extinction for this galaxy is extremely high,
$E(\mathrm{B}-\mathrm{V})_{\mathrm{GAL}}=0.342$ (see Table
\ref{sample}), in equivalency to $A_{\mathrm{FUV,NUV}}>2.5$. The
accuracy for the Galactic dust map provided by
\citet{1998ApJ...500..525S} is 16\%, corresponding to an uncertainty
of $\pm$0.055 in $E(\mathrm{B}-\mathrm{V})_{\mathrm{GAL}}$ for
NGC~6946. By inspecting the \citet{1998ApJ...500..525S} map, we
obtain a standard deviation of $\pm$0.002 around the mean value of
0.342 in $E(\mathrm{B}-\mathrm{V})_{\mathrm{GAL}}$ within the sky
coverage of NGC~6946 ($11.5\arcmin \times 9.8\arcmin$, see Table
\ref{sample}). The two contributions introduce additional
uncertainties of $\pm$0.435 in FUV magnitude and $\pm$0.440 in NUV
magnitude, and propagate to $\pm$0.005 mag in
$\mathrm{FUV}-\mathrm{NUV}$ (due to the association between the
uncertainties in FUV and NUV magnitudes) and $\pm$0.174 in IRX.
These uncertainties have a very trivial influence on the IRX-UV
relation. Nevertheless, if there is any significant variation in
extinction law at the low Galactic latitude, FUV and NUV extinctions
are likely to be biased for such a high level of
$E(\mathrm{B}-\mathrm{V})_{\mathrm{GAL}}$.

It is shown in Table \ref{contri} that, at each waveband the total
luminosity of the UV clusters accounts for 20\%$-$25\% of the
integrated luminosity of the galaxy, and a considerable amount of
integrated emission of the galaxy comes from the galactic
background; similar to NGC~5194, the center area of NGC~6946
presents no prominent feature, and the contribution of the center to
the integrated measurements of the galaxy is insignificant.
Correspondingly, the integrated galaxy lies at
$\mathrm{FUV}-\mathrm{NUV}$ = 0.60 $\pm$ 0.06 mag and IRX = 0.42
$\pm$ 0.05 which is the average position of the local background
regions, and the point of the center area populates at a remote
place. The integrated NGC~6946 follows the scatter of its subregions
and therefore presents the largest deviation in the IRX-UV location
from the starburst empirical line in our sample (Figure
\ref{IRXUV_D07}).

\begin{figure}
\centering
\vspace*{-10mm}
\includegraphics[width=\columnwidth]{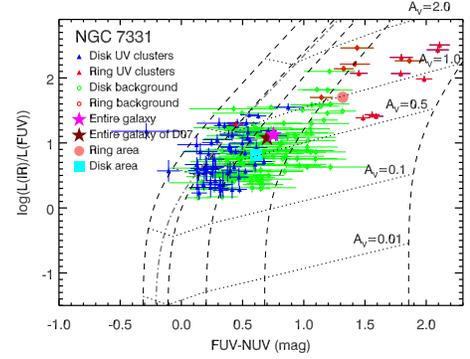}
\vspace*{-55mm}
\caption{IRX vs. $\mathrm{FUV}-\mathrm{NUV}$ of NGC~7331 for the
disk-UV clusters (blue filled triangles), the ring-UV
clusters (red filled triangles), the disk-background regions
(green open circles), the ring-background regions (red
open circles), the entire galaxy (magenta filled star), the
ring area (pink filled circle), the disk area (cyan
filled square), and the entire galaxy with fluxes cited from the D07
paper (brown filled star), with the same curves as shown in
Figure \ref{IRXUV_UV_M81} superimposed. Error bars showing the
photometric uncertainties are plotted as well.}
\label{IRXUV_UV_NGC7331}
\end{figure}

\subsection{NGC~7331}\label{7331}

In our sample, the galaxy NGC~7331 hosts a ring structure in the
central area which consists of dust and gas. This striking morphological feature is obvious from the IR imaging observations
\citep[see the right panel of Figure \ref{aper_NGC7331} in this paper,
or see][for more panchromatic pictures]{2004ApJS..154..204R}.
Stellar emission from the ring area is therefore heavily attenuated.
Figure \ref{IRXUV_UV_NGC7331} shows the IRX-UV relation for
NGC~7331, with the same curves as superimposed on the above
diagrams. The data points of the ring regions in this plot are
marked in red. One can clearly see a systematic offset between the
disk UV clusters and the disk background regions. The disk
background regions cover a redder space than the disk UV clusters of about 0.3 mag in
$\mathrm{FUV}-\mathrm{NUV}$. The
two-parameter scenario of model provides a good characterization of
this distribution: the two populations have comparable amounts of
dust attenuation; the UV clusters lie in the age range $<$ 100 Myr,
while the background regions have evolved into the older scale, with several regions at $>$ 300 Myr. Nevertheless, the same
scenario gives an inexplicable description of the locations of
ring regions. All the ring regions are located in the range of
$\mathrm{FUV}-\mathrm{NUV}>1.0$ mag in the diagram. In particular
the ring UV clusters lie in the extremely red range beyond
$\mathrm{FUV}-\mathrm{NUV}=1.5$ to over 2.0 mag \citep[this feature
has also been found in][]{2007ApJS..173..572T}. In the two-parameter
scenario of model, the dust attenuation parameter is set into
$A_\mathrm{V}>1.0$ to fit most of the ring regions, which is in
agreement with the high attenuation property of the ring area; but
the age over 100 Myr suggested by the model grid in this figure is
clearly inconsistent with the presence of strong H$\alpha$ emission
in the regions \citep[as displayed in][]{2004ApJS..154..204R,
2007ApJS..173..572T} which requires stellar population age in
principle younger than approximately 5 Myr (O-type stars dominating
environments). The error bars in this plot show that the photometric
uncertainties of the ring clusters are comparable to the typical
values of all the UV clusters ($\pm$0.07 mag in
$\mathrm{FUV}-\mathrm{NUV}$ and $\pm$0.08 in IRX), which indicates
that the photometric uncertainties are not responsible for the
deviation of the ring points, and their locations are supposed to be
physical. Again, the scenario with stellar population age as the
second parameter is confronted with a discrepancy in the
characterization for the NGC~7331 dust ring.

In Figure \ref{IRXUV_UV_NGC7331}, the point of the integrated galaxy
is located at the position of $\mathrm{FUV}-\mathrm{NUV}$ = 0.75
$\pm$ 0.06 mag and IRX = 1.12 $\pm$ 0.05, and the integrated ring
area populates at $\mathrm{FUV}-\mathrm{NUV}$ = 1.32 $\pm$ 0.06 mag
and IRX = 1.70 $\pm$ 0.05. As can be seen from Table \ref{contri},
the ring area accounts for 16\% of FUV, 26\% of NUV, and about 60\%
of the total IR luminosities of the integrated measurements of
NGC~7331. The total UV clusters contribute less than 30\% of the
integrated luminosity of the galaxy at every photometric band. This
quantification of fractional contributions offers a demonstration
of the IRX-UV location of the integrated galaxy. The point is
located at the average position of the disk regions on the
$\mathrm{FUV}-\mathrm{NUV}$ axis but between the disk and ring
regions on the IRX axis, due to the trivial UV contributions but
strong IR emission from the ring area.

\subsection{IRX-UV Relation for All The UV Clusters}

\begin{figure}
\centering
\vspace*{-10mm}
\includegraphics[width=\columnwidth]{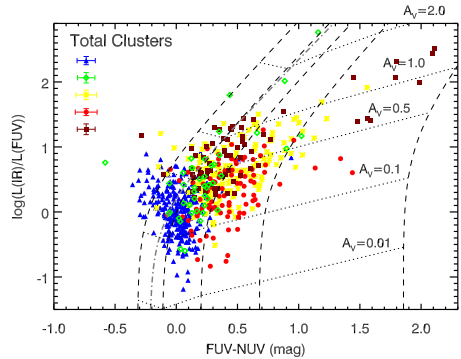}
\vspace*{-55mm}
\caption{IRX vs. $\mathrm{FUV}-\mathrm{NUV}$ for the total UV
clusters in the five galaxies: NGC~3031 (blue triangles), NGC~4536 (green diamonds),
NGC~5194 (yellow asterisks), NGC~6946 (red circles), and NGC~7331 (brown squares), with the
same curves as shown in Figure \ref{IRXUV_UV_M81} superimposed.
Error bars showing the median uncertainty for each galaxy are
plotted at the higher left corner.} 
\label{IRXUV_UV_total}
\end{figure}

In our work, we divide the aperture-extracted sources inside each
galaxy into UV clusters and local background regions, for the
purpose of separating the age parameter from the IRX-UV function.
Although in many cases UV emitters are still likely to evolve to
over 100 Myr, compared with the galactic background, the UV clusters 
are in a good position to represent young and relatively simpler
populations in galaxies. In Figure \ref{IRXUV_UV_total}, we plot
IRX-UV relation for all the UV clusters inside the galaxies in our
sample in order to examine the data distribution when the variation
in the age parameter is constrained. However, we cannot see any
uniform trend in the diagram. Instead, all the data points spread in
a wide extent, and the composite distribution of all the UV clusters
presents a scatter that is more vast than that for each individual galaxy. This
figure highlights all the discrepancies between the data
distributions and the scenario of model we have presented in the preceding
paragraphs, and proposes a challenge to the second IRX-UV parameter.
In agreement with other studies of normal star-forming galaxies
\citep{2007ApJS..173..392J} or extragalactic star-forming regions
\citep{2009ApJ...706..553B}, there has not been any consistent trend
at constant age ranges observed for different galaxies. In such
cases, the age effect becomes weak, and it is necessary to suspect
other causes of the scatter in the IRX-UV relation. In Section
\ref{disc}, we will present additional scenarios with more
complicated SFHs than the instantaneous burst we have adopted, and
discuss the possible contribution of variations in SFH to the
scatter in the IRX-UV relation.

\section{THE DEPENDANCE ON LUMINOSITY AND RADIAL DISTANCE}\label{result2}

Statistic studies of galaxies have implied a luminosity trend in the
IRX-UV distribution. They suggest that the sources with the
lowest luminosity lie at the bottom of the relational locus and the
bright objects tend to populate closer to the starburst empirical
line \citep{2007ApJS..173..185G}. This finding brings forward a
caveat that in the sampling process the preferential selection of
bright sources has the potential to omit a certain section of low
luminosities in the whole distribution of this relation. In
addition, roughly declining radial profiles of IRX and
$\mathrm{FUV}-\mathrm{NUV}$ have been found in galaxies
\citep{2007ApJS..173..524B, 2009ApJ...701.1965M}. If a radial
trend exists in galaxies, the IRX-UV locations for galactic
subregions tend to be associated with their spatial positions in
galaxies in spatially resolved studies.

In order to examine the effects of luminosity and radial distance,
we plot the IRX-UV diagrams for the subregions within the galaxies
in our sample, color-coded according to different luminosities and different
radial distances respectively. The radial distance is defined as
follows: for each galaxy, we employ four ellipses positioned at the
galactic center to define five scales of galactocentric distance;
ellipticity and position angle of the ellipses are equal to those of
the isophote of 25 mag arcsec$^{-2}$ in B band surface brightness
($D_{25}$) obtained from the NASA/IPAC Extragalactic Database; axis
lengths of the ellipses are set into 1/6, 2/6, 3/6, and 4/6 of
$D_{25}$ for the sake roughly equalizing the number of objects in
each distance range. By this approach, we obtain five ranges of radial
distance for the galactic subregions.

\begin{figure*}[!ht]
\centering
\vspace*{-10mm}
\epsscale{1.1}\plottwo{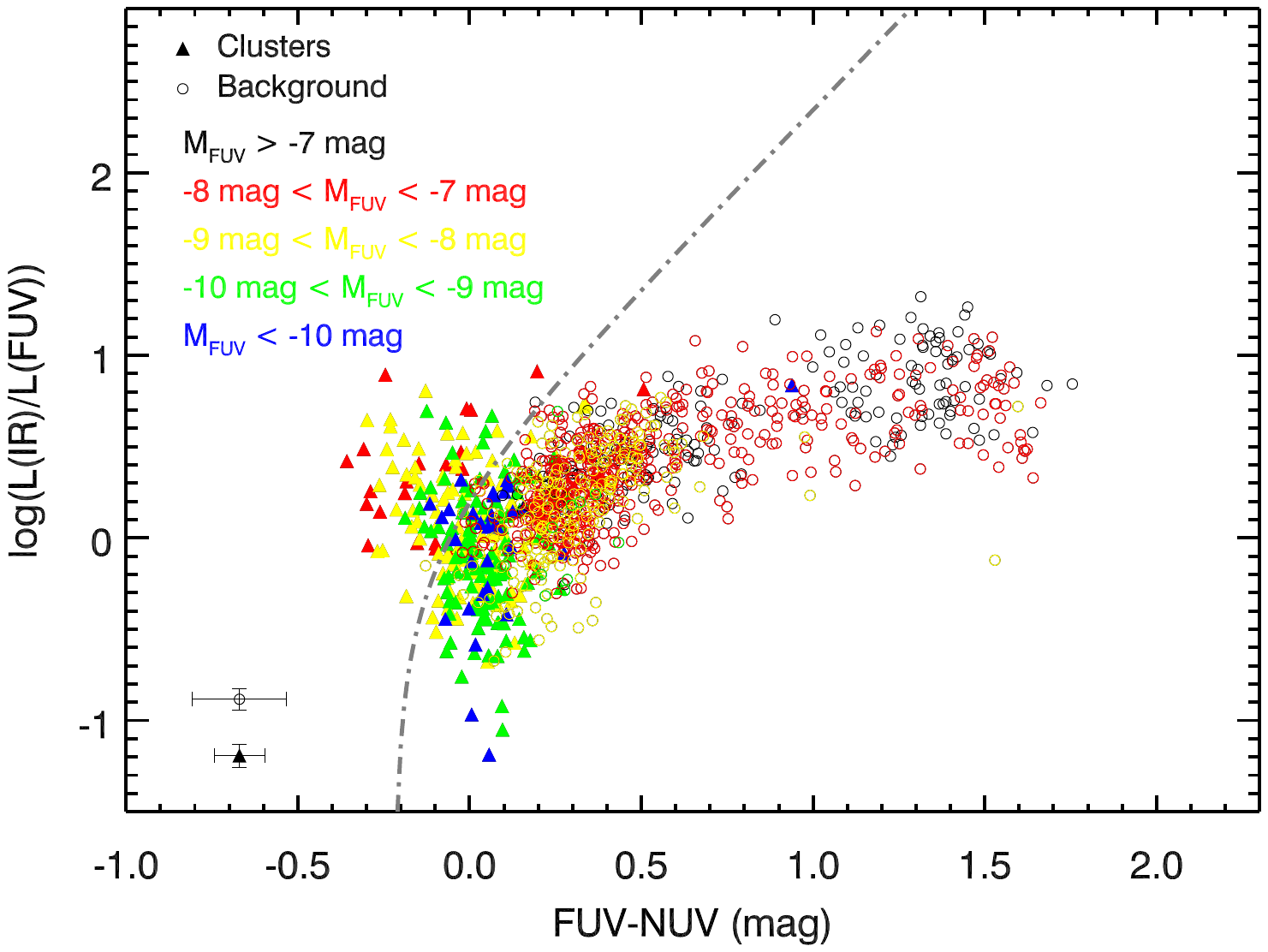}{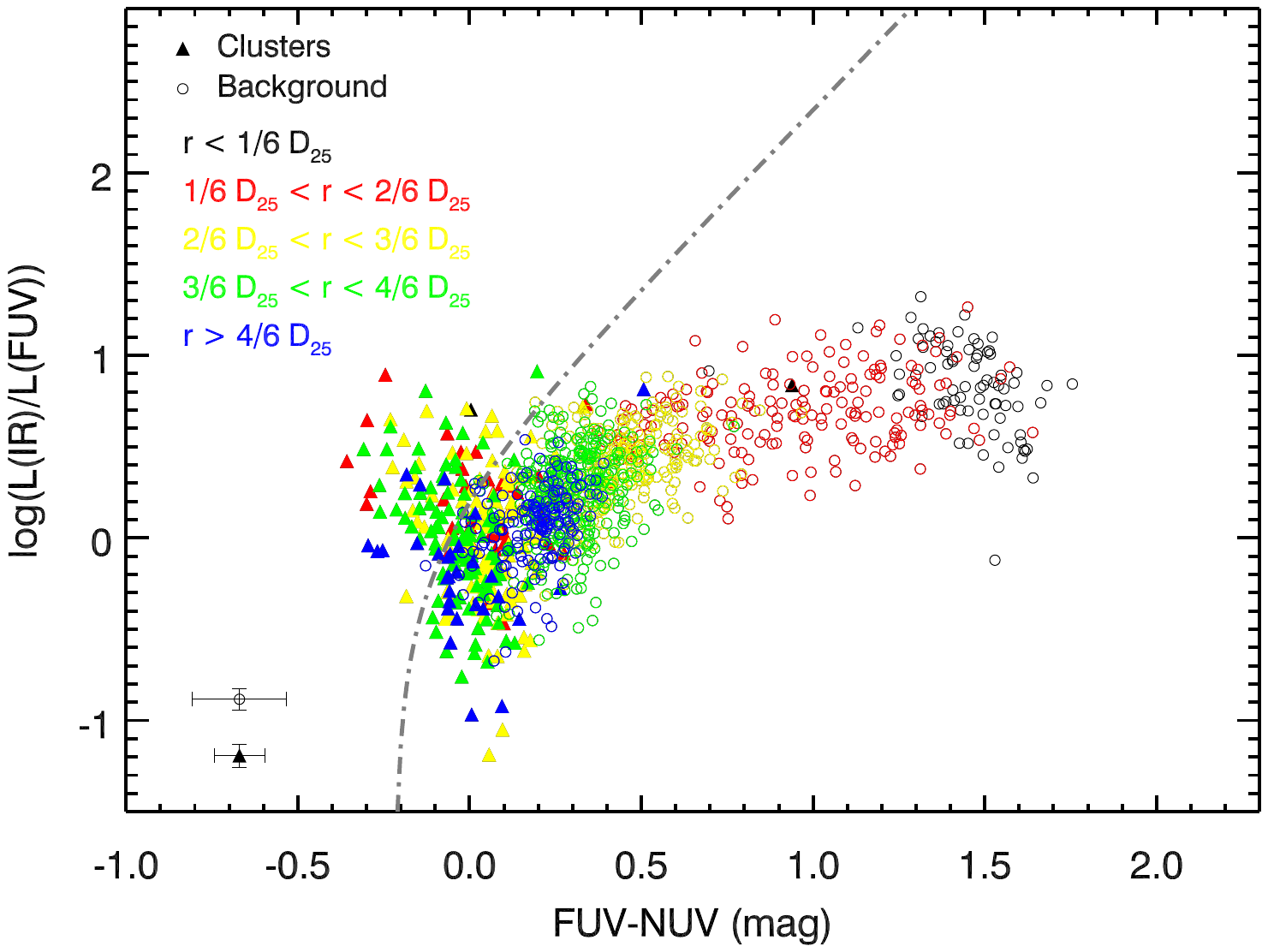}
\vspace*{-50mm}
\caption{IRX vs.
$\mathrm{FUV}-\mathrm{NUV}$ for NGC~3031 for the UV clusters
(triangles) and the local background regions (circles) color-coded
by FUV magnitude (left) and radial distance (right).
The grey line is K04 starburst empirical curve. Error bars at the
bottom left corner of each diagram show the median error for the UV
clusters (the above one) and the local background regions (the below
one).}
\label{IRXUV_2_M81}
\end{figure*}

The resulting diagrams display a weak dependence on FUV luminosity
or radial distance. The luminosity-dependent trend is contrary to the statistic studies, whereas the radial trend
coincides with results from the previous work. In our sample, the trends are
relatively obvious for NGC~3031 where there is a distinct separation
between disk and bulge. We take this galaxy as an example to show
the luminosity and radial distance trends in Figure
\ref{IRXUV_2_M81}.

The left panel of the figure presents the luminosity trend in the
IRX-UV relation. The regions with high FUV luminosity appear to have
low IRXs and blue UV colors, and the low FUV objects are located in
the range with relatively large values of IRX and
$\mathrm{FUV}-\mathrm{NUV}$. The inconsistency to the studies of
integrated galaxies is due to different natures of the targets in
the respective work. In the integrated results, most of the
FUV-brighter objects close to the starburst empirical relation are
starburst or star-forming galaxies, while dwarf and irregular
galaxies with faint luminosities and low IRX make up the bottom part
of the diagram due to their low mass and dust-poor state
\citep{2007ApJS..173..185G, 2009ApJ...703..517D,
2009ApJ...706..599L}. In our work, the regions with low FUV
luminosity are defined as dust-rich sources or evolved stellar
populations with large infrared-to-ultraviolet ratios and red
colors, and therefore located at the top part of the diagram;
whereas the strong UV emission arises from dust-poor regions and
young clusters which lead to the declining trends of the IRX and
$\mathrm{FUV}-\mathrm{NUV}$. The radial dependance of the IRX-UV
relation is shown at the right panel of the figure. It is apparent
that the both parameters in the relation decrease at larger radii,
in accordance with the studies of galactic radial profiles
\citep{2007ApJS..173..524B, 2009ApJ...701.1965M}. For NGC~3031, the
apparent trend is better indicated by the disk-bulge separation: the
central bulge contains the oldest populations in the galaxy with
intrinsically low FUV luminosity and short radial distance.

\section{TWO-DIMENSIONAL MAPS OF FUV $-$ NUV AND IRX}\label{2-D map}

\begin{figure*}
\centering \epsscale{0.9}\plottwo{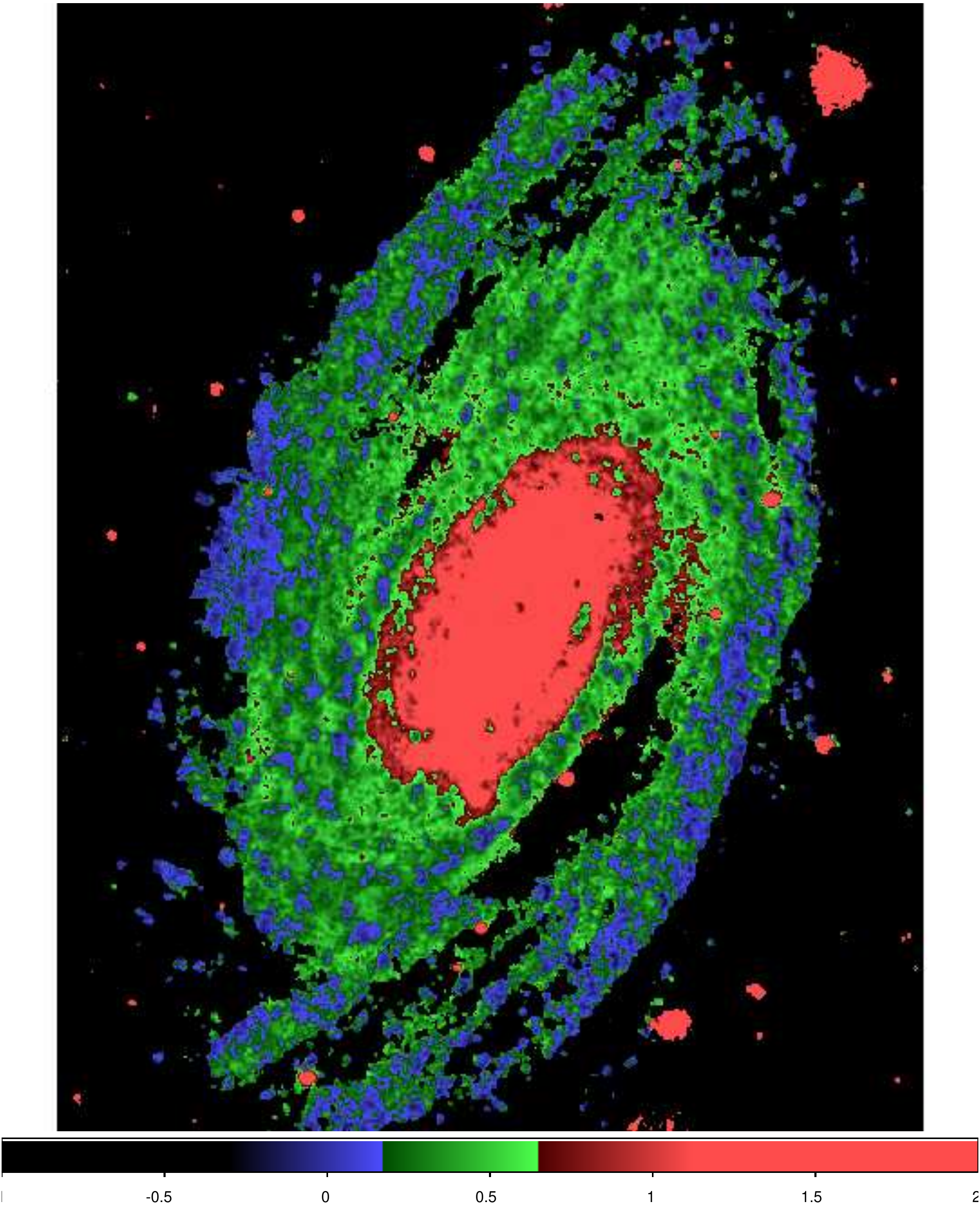}{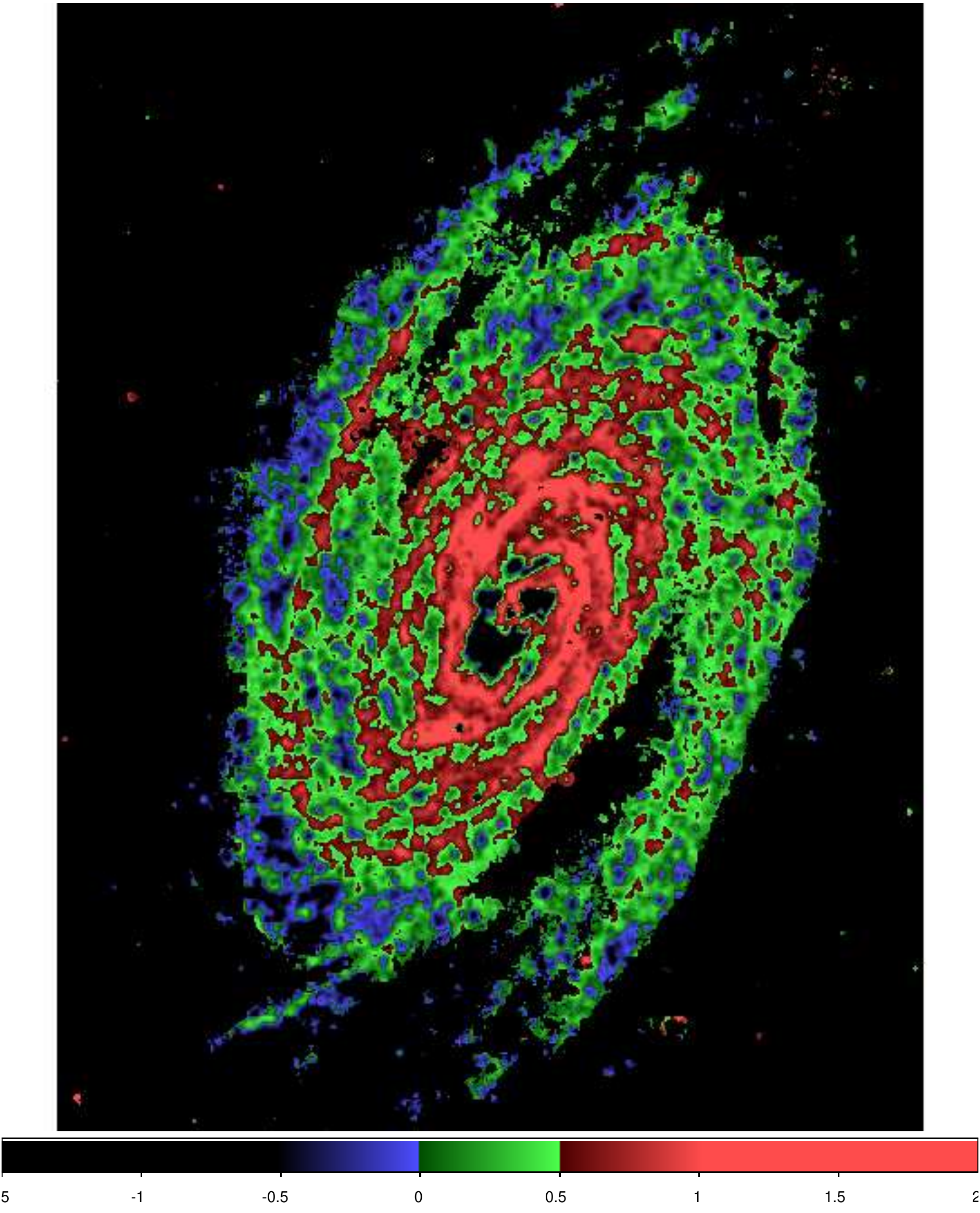}
\caption{2-D maps of
$\mathrm{FUV}-\mathrm{NUV}$ (left) and IRX (right) for
NGC~3031. Color scales of $\mathrm{FUV}-\mathrm{NUV}$ and IRX are
shown on the bottom bar at each panel. The size of the both maps is
$15.6'\times20.4'$. North is up and east is to the
left.}\label{maps_M81}
\end{figure*}

\begin{figure*}
\centering
\epsscale{0.9}\plottwo{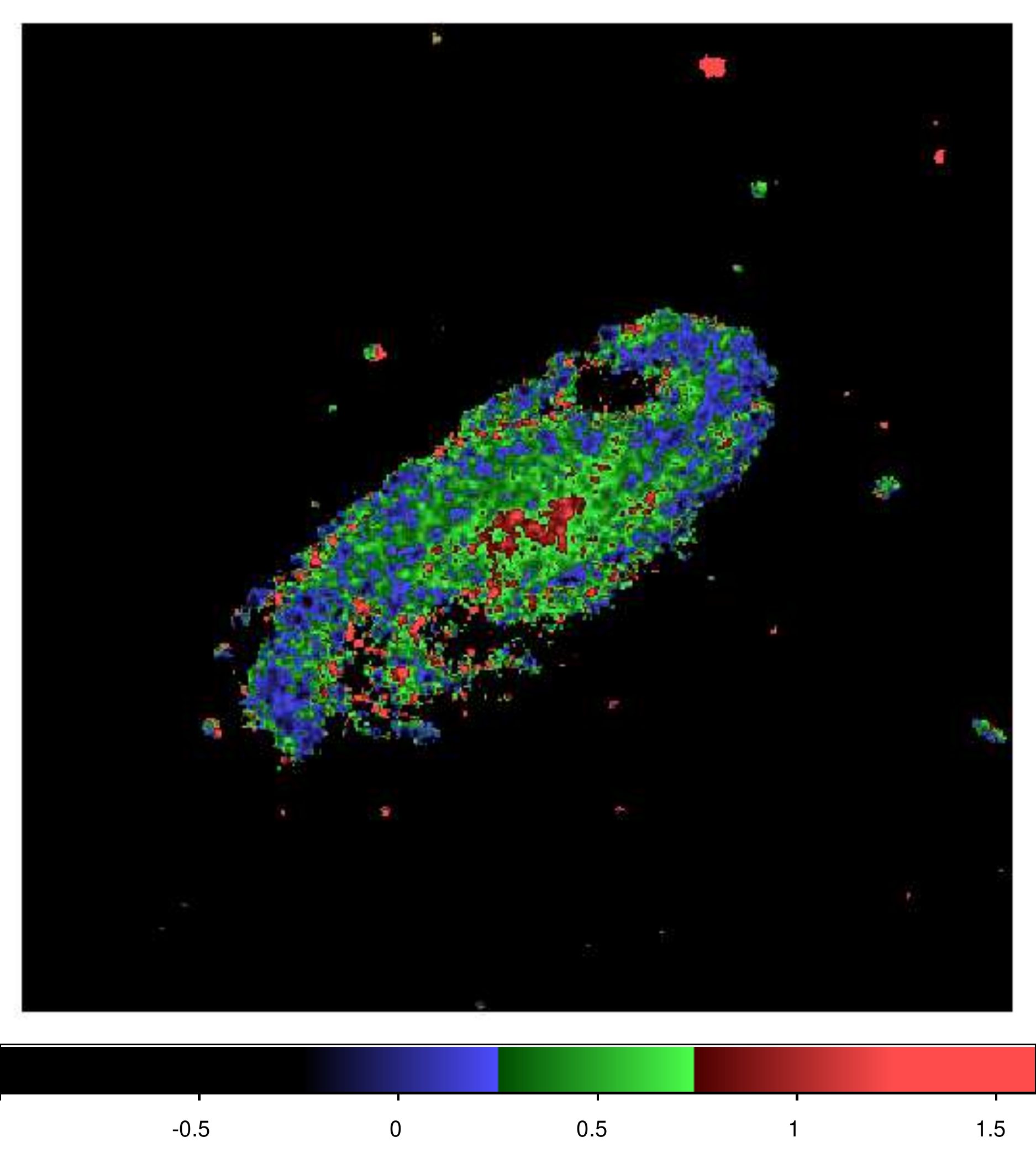}{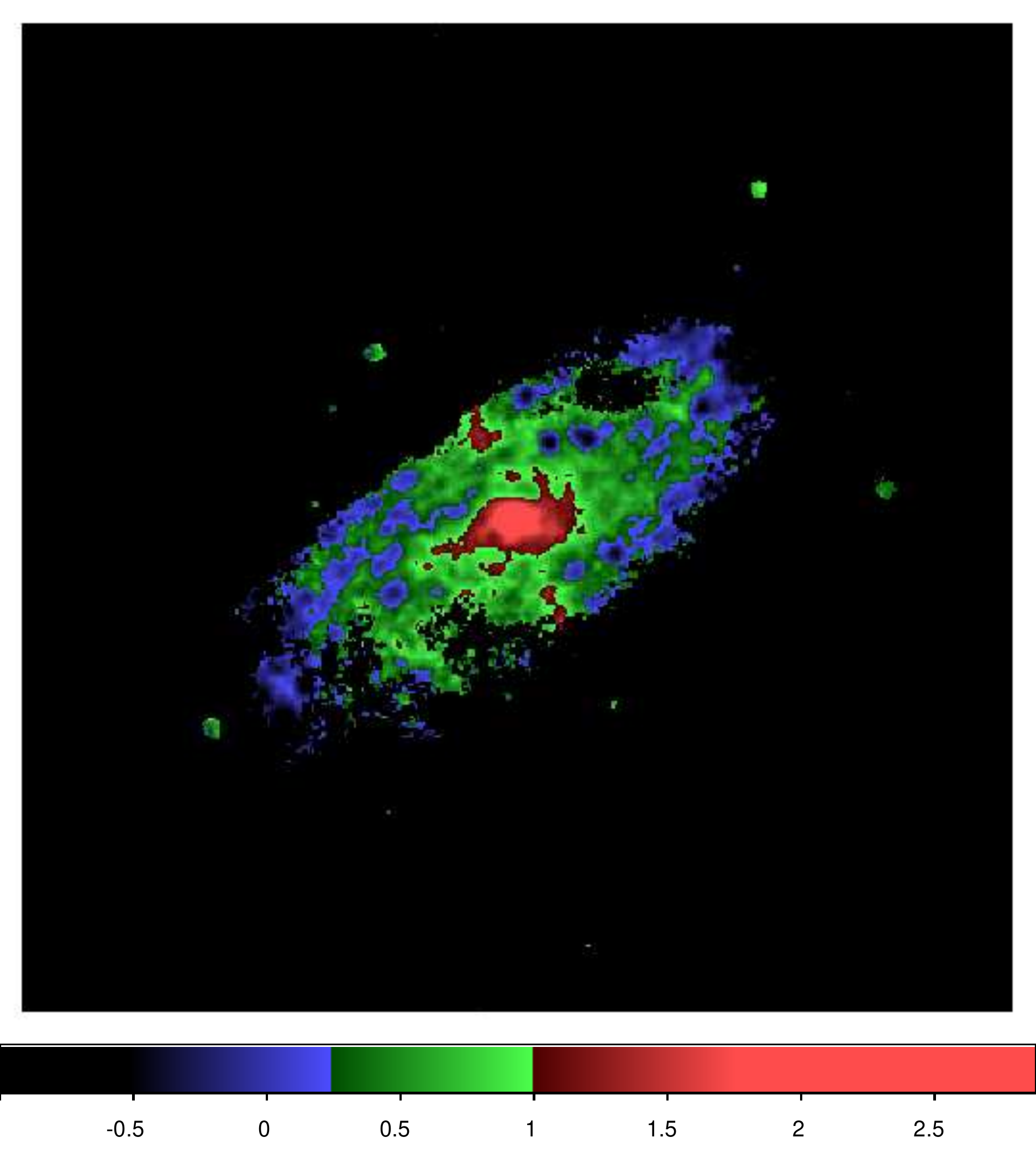}
\caption{2-D maps of
$\mathrm{FUV}-\mathrm{NUV}$ (left) and IRX (right) for
NGC~4536. Color scales of $\mathrm{FUV}-\mathrm{NUV}$ and IRX are
shown on the bottom bar at each panel. The size of the both maps is
$11.4'\times11.4'$. North is up and east is to the
left.}\label{maps_NGC4536}
\end{figure*}

\begin{figure*}
\centering \epsscale{0.9}\plottwo{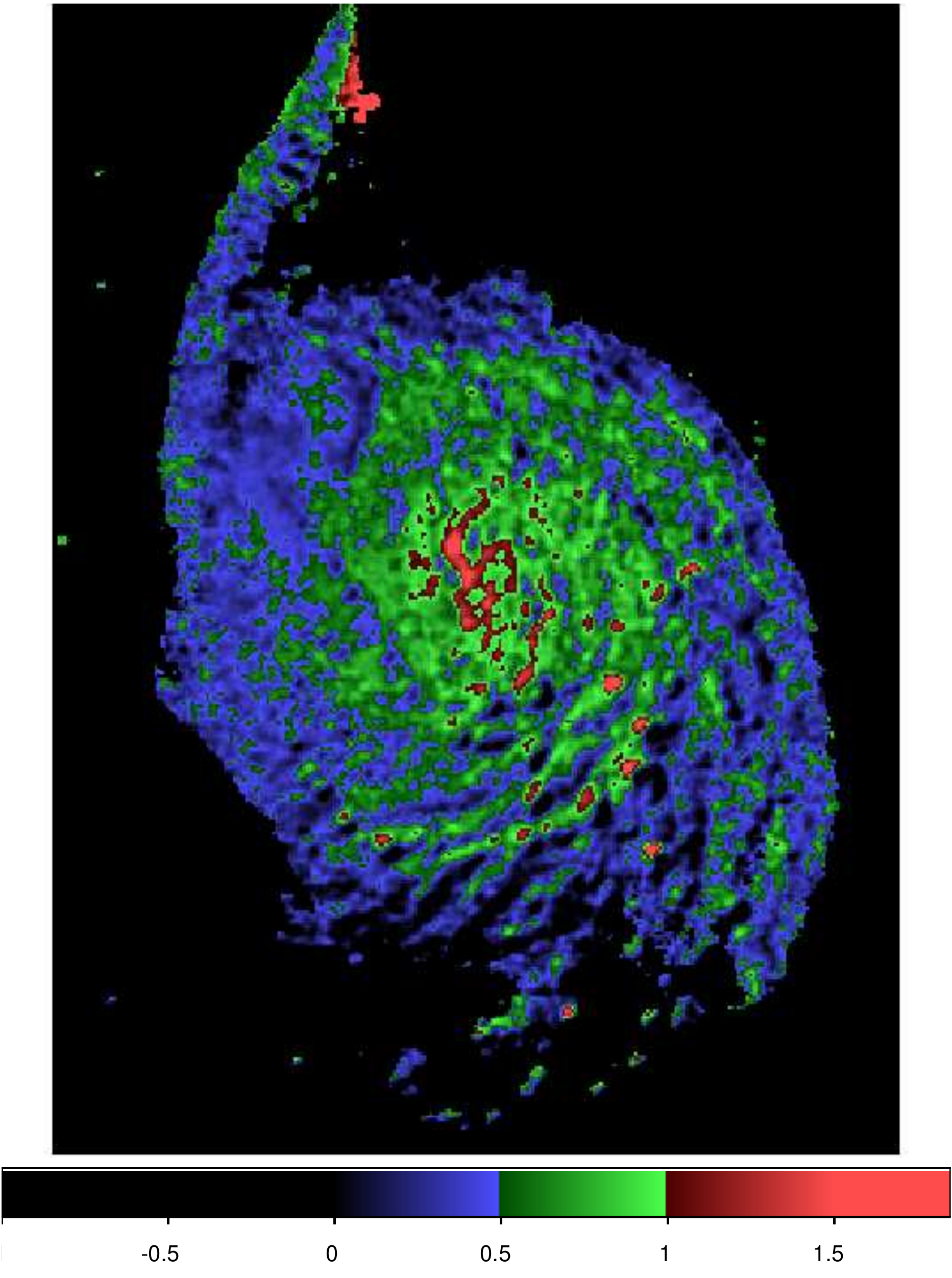}{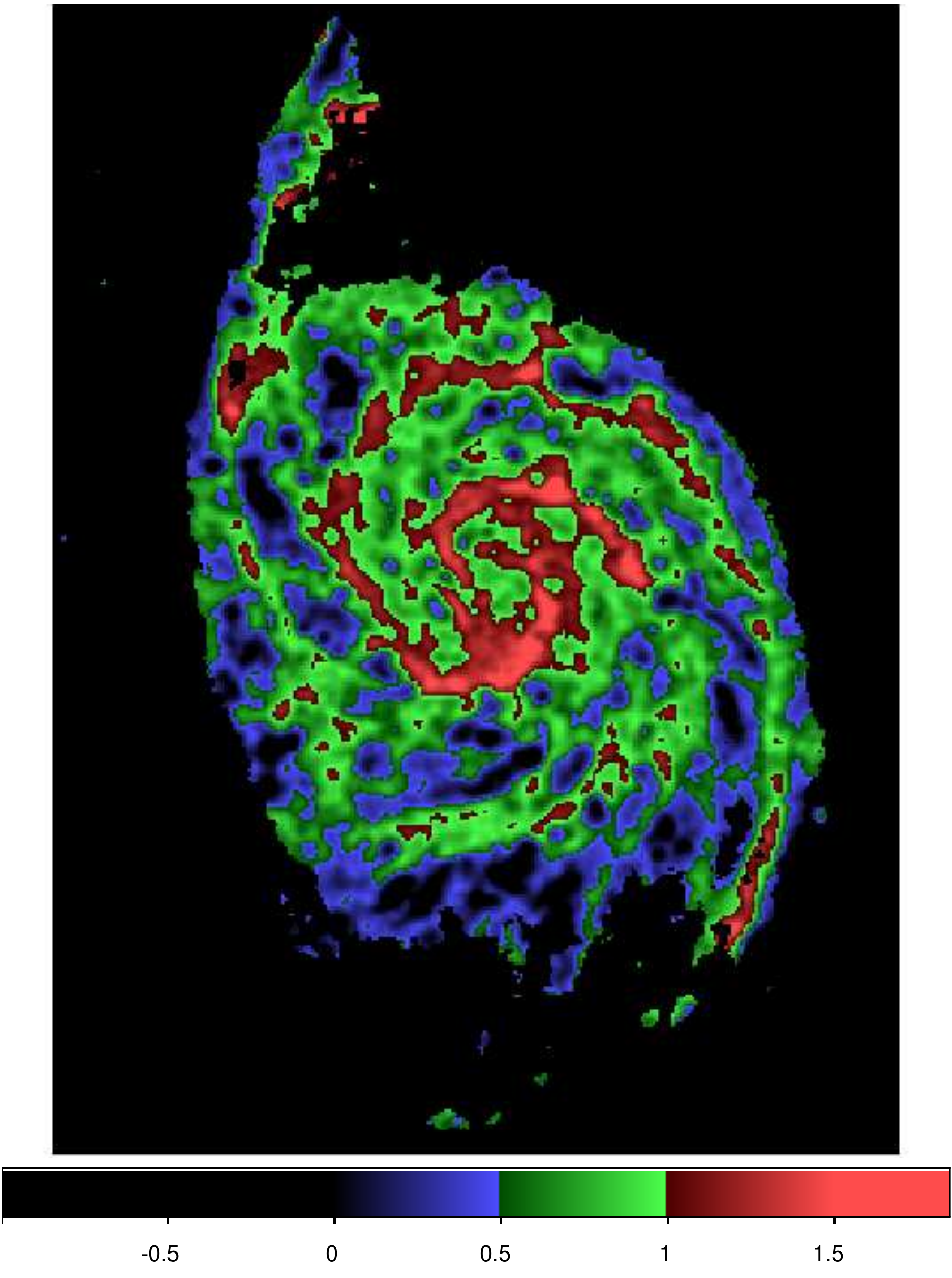}
\caption{2-D maps of
$\mathrm{FUV}-\mathrm{NUV}$ (left) and IRX (right) for
NGC~5194. Color scales of $\mathrm{FUV}-\mathrm{NUV}$ and IRX are
shown on the bottom bar at each panel. The size of the both maps is
$8.0'\times8.0'$. North is up and east is to the
left.}\label{maps_M51}
\end{figure*}

\begin{figure*}
\centering
\epsscale{0.9}\plottwo{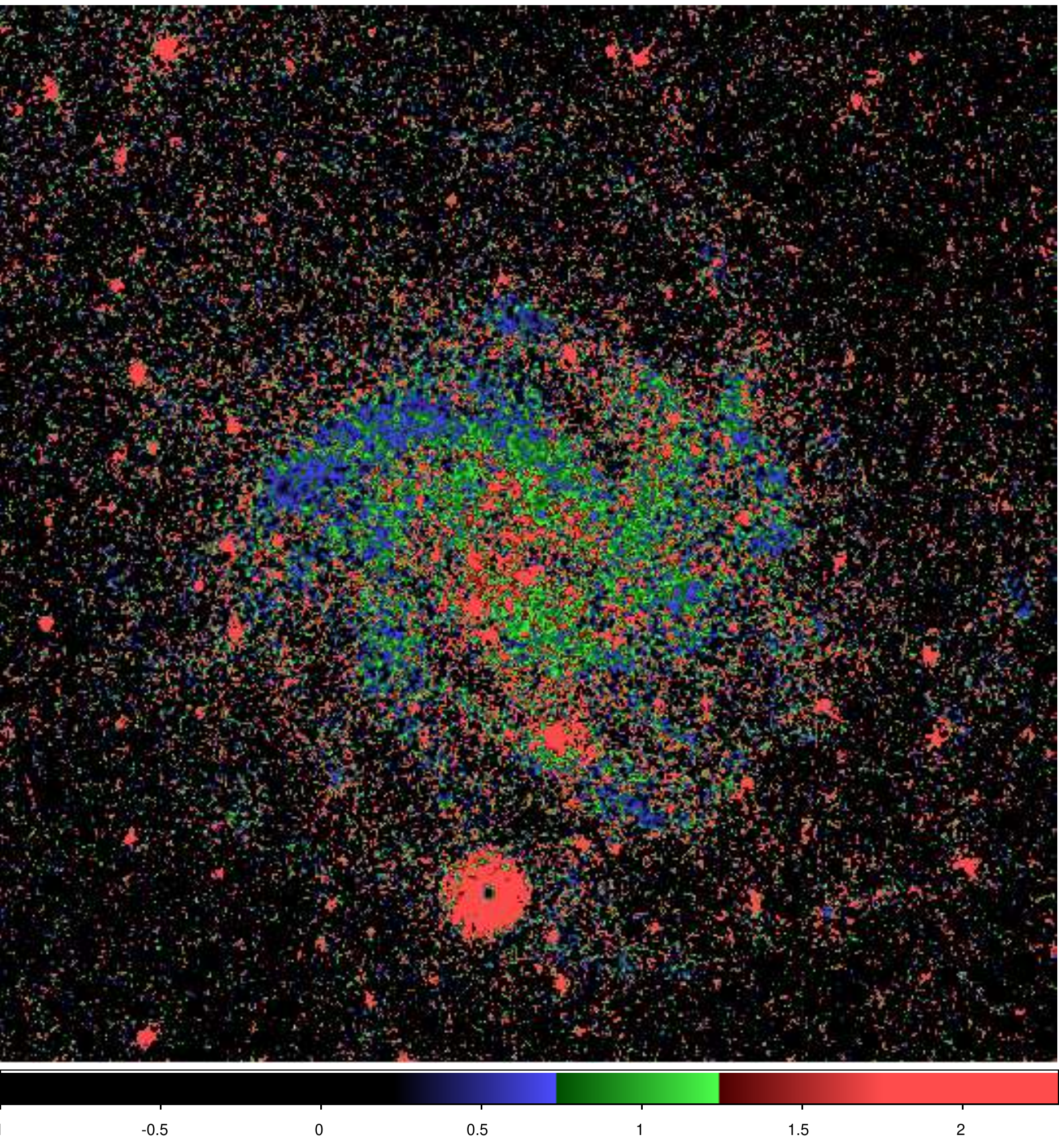}{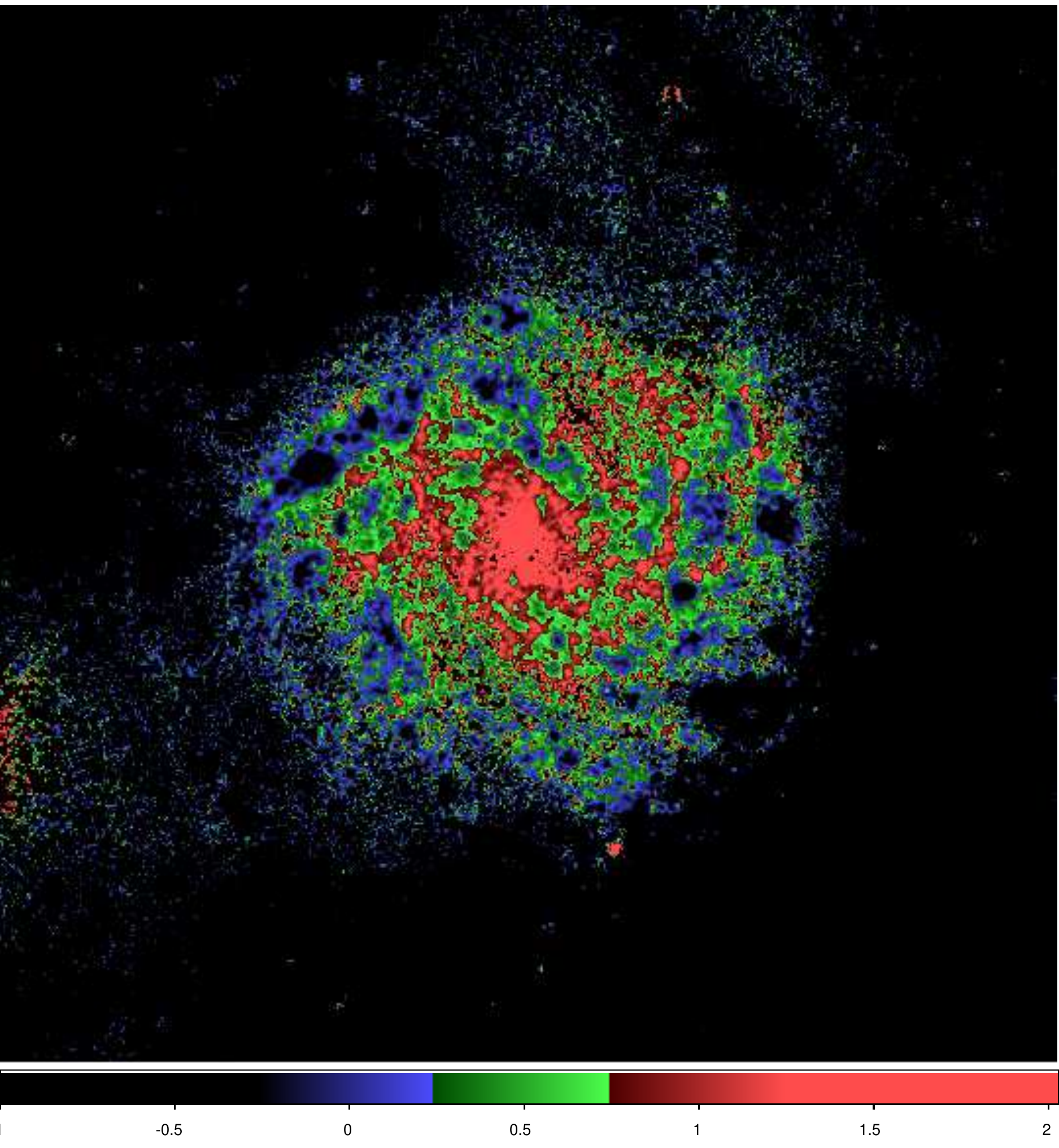}
\caption{2-D maps of
$\mathrm{FUV}-\mathrm{NUV}$ (left) and IRX (right) for
NGC~6946. Color scales of $\mathrm{FUV}-\mathrm{NUV}$ and IRX are
shown on the bottom bar at each panel. The size of the both maps is
$17.3'\times17.3'$. North is up and east is to the
left.}\label{maps_NGC6946}
\end{figure*}

\begin{figure}
\centering
\epsscale{1.0}\plottwo{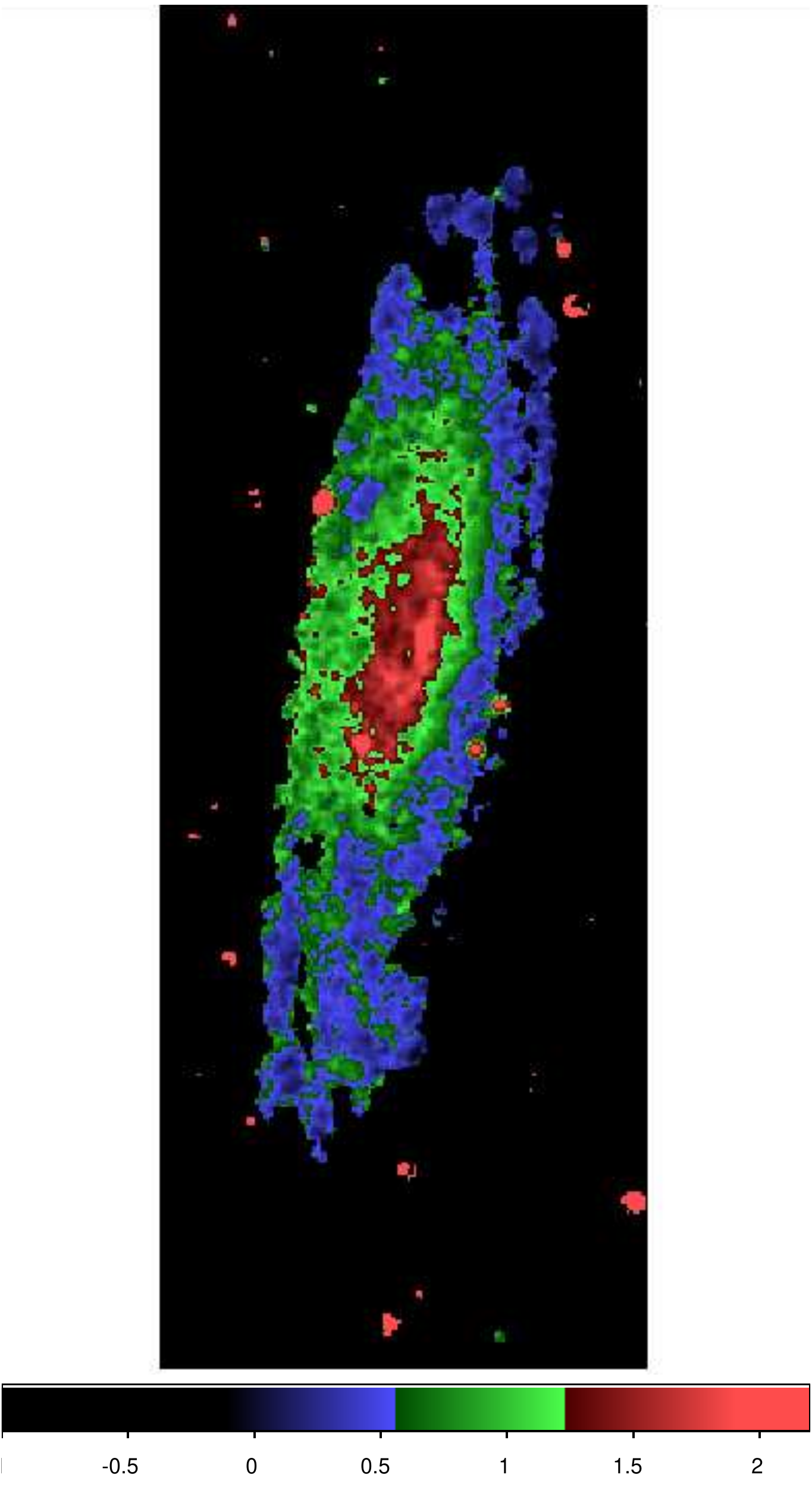}{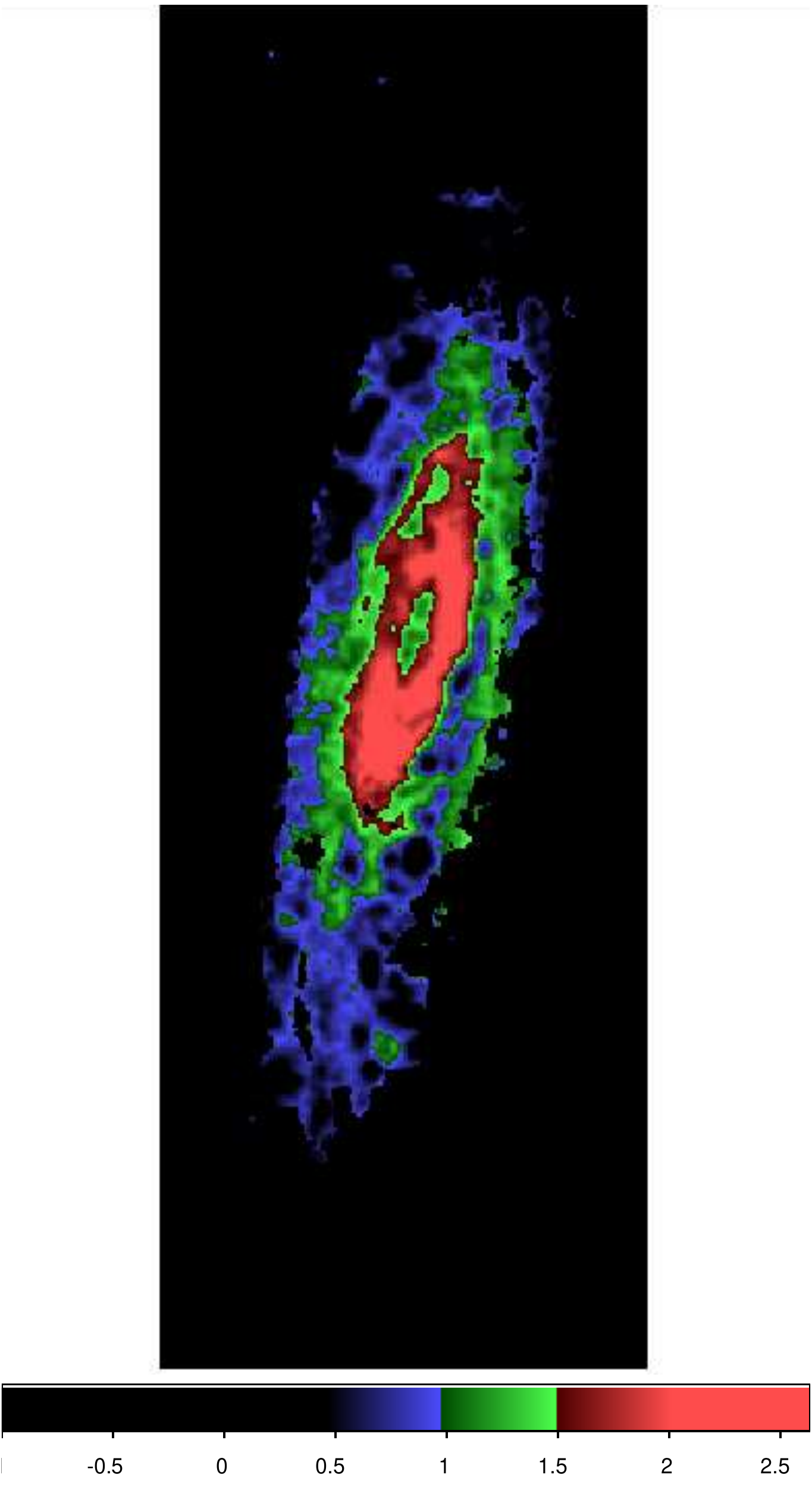}
\caption{2-D maps of
$\mathrm{FUV}-\mathrm{NUV}$ (left) and IRX (right) for
NGC~7331. Color scales of $\mathrm{FUV}-\mathrm{NUV}$ and IRX are
shown on the bottom bar at each panel. The size of the both maps is
$4.5'\times12.7'$. North is up and east is to the
left.}\label{maps_NGC7331}
\end{figure}

In order to scan the spatial distributions of
$\mathrm{FUV}-\mathrm{NUV}$ color and IRX with clear circumstance,
we plot the two-dimensional maps of the two parameters for each
galaxy in Figures \ref{maps_M81}$-$\ref{maps_NGC7331}. In each
figure, the highest values of the $\mathrm{FUV}-\mathrm{NUV}$ color 
is in the center, and at larger radii the value decreases finally falling to the minimum at the galactic edge. The IRX
presents the similar descending gradient to the UV color but a
differentiation in pattern. There is no obvious fluctuation of the
$\mathrm{FUV}-\mathrm{NUV}$ along galactic radii, and the color
change along the radius is quite smooth; whereas on the IRX maps we
can clearly see spiral arm-like structures, and in addition, the
inner parts of spiral arms seem to exhibit higher IRXs than the
outer parts in some cases. This feature is more obvious in
the face-on galaxies NGC~3031, NGC~5194, and NGC~6946. Although
the edge-on galaxies fail to illuminate the same details, in our
sample NGC~4536 still exhibits the central clump and NGC~7331
displays the ring structure in centers on their respective IRX maps.
The diversity in the spatial distributions between UV color and IRX
can be elucidated with the different natures between the two
parameters. The infrared-to-ultraviolet ratio is in tight
correlation to dust attenuation in general cases compared to the UV
color, and thus the IRX maps are believed to take better pictures of
dust attenuation distributions in galaxies. On the contrary, the UV
color is sensitive to not only one single parameter, and the
degeneracy of parameters in UV color is very likely to erase any
character of single factor. The smooth gradients in the
$\mathrm{FUV}-\mathrm{NUV}$ maps are a reflection of the
compromise. The mismatch in spatial positions for the two parameters
is the visual presentation of the scatter in the IRX-UV diagram.

\section{DISCUSSION}\label{disc}

The main goal of this work is to have a better understanding of the
two parameters (dust attenuation and stellar population age) in the
IRX-UV function. In Section \ref{result1}, we present the spatially
resolved results of the IRX-UV diagrams. We can see the age effects
appearing as the displacements between the locations of the UV
clusters and the local background regions inside galaxies in these
diagrams, and the deviations of the age parameter in the
interpretations of the IRX-UV distributions. The two-parameter
scenario adopted for characterizing the observational data in
Section \ref{result1} assumes simple stellar populations with an
instantaneous burst. This assumption is an approximation of stellar
populations contained in galactic subregions. As a matter of fact,
although subregions inside galaxies are simpler in SFH than
integrated galaxies, stellar populations still tend to be born
beyond the instantaneous burst on a scale of several hundreds
parsecs, which has the potential to oversimplify the stellar
populations and introduce inaccuracies in the resulting descriptions
of age for the measured regions. In this section, we will propose a
series of composite stellar populations for the modeled scenarios,
and discuss potential influences of the complexities in SFH on the
IRX-UV properties.

\begin{figure}
\centering
\vspace*{-10mm}
\includegraphics[width=\columnwidth]{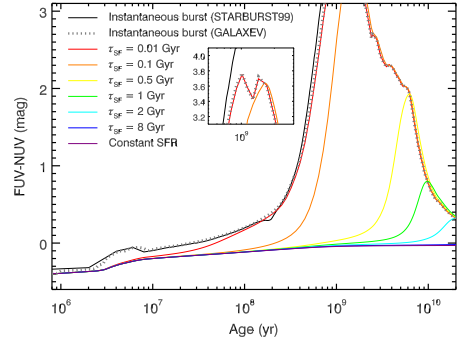}
\vspace*{-55mm}
\caption{$\mathrm{FUV}-\mathrm{NUV}$ color as a function of stellar
population age for different SFHs: simple stellar populations with
instantaneous burst, composite stellar populations with exponential
decreases in SFR, and constant SFR. The instantaneous burst
functions are derived from two models of stellar population
synthesis: STARBURST99 (black solid line) and GALAXEV
(gray dotted line), for comparison. The exponentially
decreasing SFRs with time constants of $\tau_\mathrm{SF}$ = 0.01
(red solid line), 0.1 (orange solid line), 0.5
(yellow solid line), 1 (green solid line), 2
(cyan solid line), and 8 Gyr (blue solid line), and
the constant SFR of 1 $M_{\odot}~yr^{-1}$ (purple solid line)
are derived from GALAXEV synthesis model. The simple and composite
stellar populations are modeled with solar metallicity ($Z=0.02$).
The middle panel is plotted to show the reversing points on the
modeled tracks with $\tau_\mathrm{SF} \leq 0.1$
Gyr.}\label{color_vs_age}
\end{figure}

Spectra of the composite stellar populations are constructed via the
GALAXEV library of evolutionary population synthesis
\citep{2003MNRAS.344.1000B}, on an assumption of exponentially
decreasing SFR ($\sim e^{-t/\tau_\mathrm{SF}}$, where $t$ is stellar
population age and $\tau_\mathrm{SF}$ is constant of star formation
timescale), with the \citet{2003PASP..115..763C} IMF \footnote{The
GALAXEV spectral library adopts the Chabrier IMF in stellar
population synthesis, while the STARBURST99 spectral library offers
the Kroupa IMF. Both IMFs are essentially identical for stellar
mass $\geq 1 M_\odot$. The difference between the two IMFs exists in
descriptions of low-mass stars and brown dwarfs, and has no
effective influence on UV stellar emission studied in this work.}
and solar metallicity ($Z=0.02$). Figure \ref{color_vs_age} shows
intrinsic $\mathrm{FUV}-\mathrm{NUV}$ as a function of stellar
population age in various SFH scenarios: simple stellar populations
with an instantaneous burst, composite stellar populations with six
types of exponential decreases in SFR ($\tau_\mathrm{SF}$ = 0.01,
0.1, 0.5, 1, 2, and 8 Gyr), and constant SFR of 1 $M_\odot~yr^{-1}$.
We produce two scenarios of an instantaneous burst in this figure
from STARBURST99 and GALAXEV respectively for the purpose of
comparison between the two libraries. The exponentially decreasing
and constant SFRs are derived from the GALAXEV library. As clearly
illustrated in this figure, stellar populations born with an
instantaneous burst present the most tremendous reddening of UV
color with increasing age, in particular at age $>$200 Myr:
$\mathrm{FUV}-\mathrm{NUV}$ extends from 0.3 to $\sim$3.0 mag within
an age range of 200$-$700 Myr. The reddening evolution of UV color
tends to fade in the scenarios of composite stellar populations, and
it is apparent to see at longer star formation time that
$\mathrm{FUV}-\mathrm{NUV}$ appears to be more insensitive to age.
The exponentially decreasing SFR with $\tau_\mathrm{SF}$ = 0.01 Gyr
has a color evolution comparable to the instantaneous burst; at
$\tau_\mathrm{SF}$ = 0.1 Gyr, the reddening of UV color becomes
intensive after 500 Myr age; for longer star formation time,
$\mathrm{FUV}-\mathrm{NUV}$ retains a constant value $\sim$0.0 mag
until stellar populations evolve to $\sim$2 Gyr at
$\tau_\mathrm{SF}$ = 0.5 Gyr, $\sim$5 Gyr at $\tau_\mathrm{SF}$ = 1
Gyr, and $\sim$10 Gyr at $\tau_\mathrm{SF}$ = 2 Gyr; stellar
populations with $\tau_\mathrm{SF}$ = 8 Gyr and the constant SFR
produce identical evolutionary tracks of UV color, where
$\mathrm{FUV}-\mathrm{NUV}$ $\sim$ 0.0 mag during the lifetime up to
20 Gyr. Complex SFHs introduce the degeneracy of stellar population
age and star formation timescale in $\mathrm{FUV}-\mathrm{NUV}$:
larger $\tau_\mathrm{SF}$ means more extended SFH and results in
bluer color than smaller $\tau_\mathrm{SF}$ at constant age.

\begin{figure}
\centering
\vspace*{-10mm}
\includegraphics[width=\columnwidth]{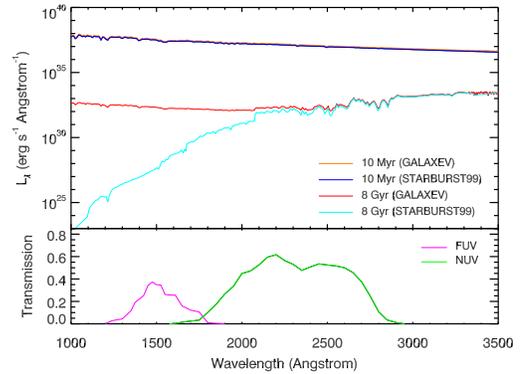}
\vspace*{-55mm}
\caption{Top: Modeled spectra in UV wavelength range of stellar
populations with instantaneous burst, $10^5 M_{\odot}$, and solar
metallicity ($Z=0.02$). The spectra are color-coded by stellar
population age and relevant models: 10 Myr from GALAXEV model
(orange), 10 Myr from STARBURST99 model (blue), 8 Gyr
from GALAXEV model (red), and 8 Gyr from STARBURST99 model
(cyan). Bottom: Filter transmission curves of two
\emph{GALEX} bandpasses: FUV (magenta) and NUV
(green).}\label{comp_model}
\end{figure}

It is worth noting in Figure \ref{color_vs_age} that, there is a
reversing point on each of the evolutionary tracks produced with the
GALAXEV library except the SFHs of $\tau_\mathrm{SF}$ = 8 Gyr and
the constant SFR, which suggests an upper limit for UV color at a
certain age, and after this stage the color is supposed to
experience a kind of blueing evolution. This feature is different
from the STARBURST99 product where $\mathrm{FUV}-\mathrm{NUV}$ is on
the monotonic increase during the whole lifetime of stellar
populations, by comparison between the two models for the same
instantaneous burst in this figure. It is also displayed that the
reversing points on the GALAXEV tracks emerge in later periods and
the upper limits appear at lower values if stellar populations
evolve with larger $\tau_\mathrm{SF}$. As shown in this figure, the
SFHs of an instantaneous burst and $\tau_\mathrm{SF}<0.1$ Gyr yield
the reddest colors over 3.6 mag during 1$-$2 Gyr; for
$\tau_\mathrm{SF}$ = 2 Gyr, the maximum in
$\mathrm{FUV}-\mathrm{NUV}$ drops to $\sim$0.3 mag, and the time
falls to $\sim$20 Gyr; and between them, the reversing point for
longer star formation time appears to locate on the track for lower
$\tau_\mathrm{SF}$, and after the reversing point the both SFHs tend
to share a common evolutionary track. The presence of the blueing
evolution is ascribed to the additional account of FUV emission from
late-type post-asymptotic-giant-branch (post-AGB) stars in the GALAXEV library
\citep{2003MNRAS.344.1000B}. Figure \ref{comp_model} shows a
comparison between UV spectral energy
distributions (SEDs) derived respectively from the STARBURST99
and GALAXEV libraries for an instantaneous burst at two speciﬁc ages
as an example to illustrate the difference in spectral shape. The
contribution of the post-AGB stars manifests as the constant
spectrum with an even rising trend shortwards from 2000
$\mathrm{Ang}$ wavelength derived from the GALAXEV library for 8 Gyr
age; whereas the STARBURST99 spectrum for this age presents a steep
decline in the same wavelength range.

\begin{figure*}
\centering
\vspace*{-10mm}
\includegraphics[width=1.8\columnwidth]{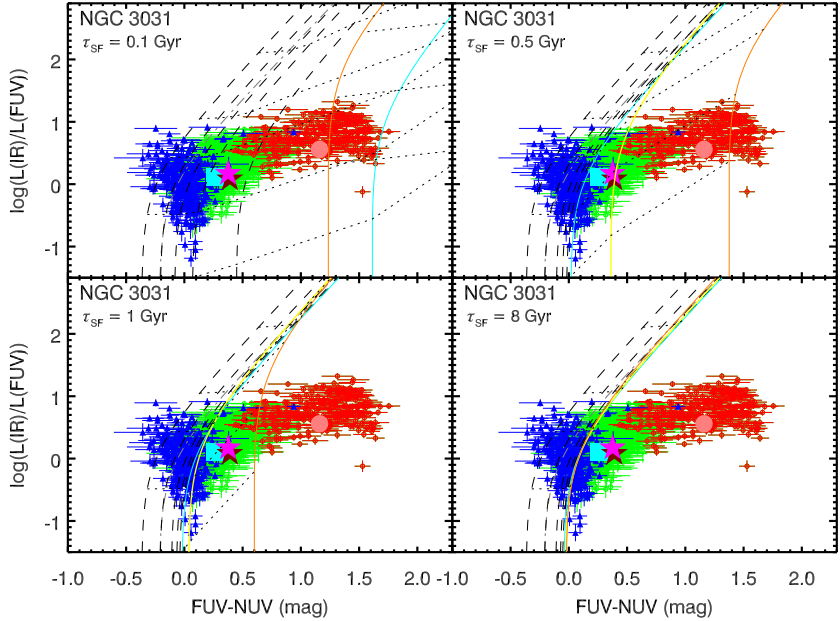}
\vspace*{-90mm}
\caption{IRX vs. $\mathrm{FUV}-\mathrm{NUV}$ for NGC~3031, and
superimposed is the grid modeled with exponentially decreasing SFRs
with time constants of $\tau_\mathrm{SF}$ = 0.1 (top left
panel), 0.5 (top right panel), 1 (bottom left panel),
and 8 Gyr (bottom right panel). Symbols are the same with
those in Figure \ref{IRXUV_UV_M81}. Black dashed lines describe the
model curves sampled with five ages: 2, 8, 100, 300, and 500 Myr,
from left to right on the horizontal axis, corresponding to those in
Figures \ref{IRXUV_UV_M81}$-$\ref{IRXUV_UV_NGC7331}. Solid lines
indicate older ages and are color-coded by 800 Myr (cyan), 3
Gyr (yellow), and 8 Gyr (orange). Dotted lines connect
the points of five constant amounts of dust attenuation
($A_\mathrm{V}$ = 0.01, 0.1, 0.5, 1.0, and 2.0) on the model curves
of different ages, corresponding to those in Figures
\ref{IRXUV_UV_M81}$-$\ref{IRXUV_UV_NGC7331}. Grey dot-dashed line is
K04 starburst curve. Error bars showing the photometric
uncertainties are plotted as well.}\label{IRXUV_UV_M81_SFH}
\end{figure*}

In Figure \ref{IRXUV_UV_M81_SFH}, we superimpose grids reproduced
with composite stellar populations on the IRX-UV diagrams for
NGC~3031 as plotted in Figure \ref{IRXUV_UV_M81}, and four types of
exponential decreases in SFR ($\tau_\mathrm{SF}$ = 0.1, 0.5, 1, and
8 Gyr) are shown respectively in each panel. The modeled curves with
constant ages are displayed in parallel with each other in the
figure, and with increasing ages the equivalent amounts of dust
attenuation tend to stand at moderately higher levels of IRX.
Compared with the SFH with an instantaneous burst, these curves for
composite stellar populations are confined in narrower
$\mathrm{FUV}-\mathrm{NUV}$ spaces, and the IRX-UV functions with
different ages tend to approach each other with increasing
$\tau_\mathrm{SF}$. For instance, in the IRX-UV planes, an age
interval between 2 and 500 Myr for $\tau_\mathrm{SF} = 0.1$ Gyr
introduces a difference of $\sim$0.7 mag in
$\mathrm{FUV}-\mathrm{NUV}$, while in the same age scale but for
$\tau_\mathrm{SF} \geq 0.5$ Gyr, the color differences diminish to
less than 0.4 mag. In addition to the five certain ages addressed in
Section \ref{result1} (Figures
\ref{IRXUV_UV_M81}$-$\ref{IRXUV_UV_total}), we further sample three
older ages in Figure \ref{IRXUV_UV_M81_SFH}: 800 Myr, 3 Gyr, and 8
Gyr. In the panel of $\tau_\mathrm{SF} = 0.1$ Gyr, the curve with
the age of 800 Myr lies in a range where $\mathrm{FUV}-\mathrm{NUV}
> 1.5$ mag;\footnote{In this panel, the curve with the age of 8 Gyr
suffers from the color reversing as we have presented in the above
paragraph, and due to this factor this curve appears on a bluer
location than that of 800 Myr.} while for $\tau_\mathrm{SF} \geq
0.5$ Gyr, the same regimes in the IRX-UV diagrams belong to stellar
populations evolving over 8 Gyr; when $\tau_\mathrm{SF}$ increases
to 8 Gyr, ages ranging from 100 Myr to 8 Gyr are not distinguishable
in the IRX-UV planes, and stellar populations in this timescale have
intrinsic colors $\mathrm{FUV}-\mathrm{NUV} \sim 0.0$ mag.

The model with composite stellar populations manifests
substantial disparities in description of the same observational
data with different SFH scenarios. In our work, the UV clusters
inside the galaxies are measured with the subtraction of underlying
diffuse emission (see Section \ref{measure}), which enables
the extraction stellar populations with an instantaneous burst or short
star formation timescale, and $\tau_\mathrm{SF} > 0.1$ Gyr is not
considered to be appropriate for the UV clusters; by contrast,
the local background populations are supposed to experience
longer-term star formation in SFH. In this situation, we expect
to adopt diverse SFH scenarios to characterize the two populations.

The IRX-UV diagram for NGC~3031 shows a clear separation between
the UV clusters, the disk background regions, and the bulge
background regions (Figure \ref{IRXUV_UV_M81}). As we have noted
in Section \ref{3031}, a considerable number of the disk background
regions are categorized in the same age range for the UV
clusters in the scenario of simple stellar populations, which is one
of the discrepancies between the observational data and the model
scenario in this paper. To address this problem, the model of composite
stellar populations offers a suggestion from the viewpoint of SFH,
where the local background regions are characterized as older
stellar populations with increasing star formation timescales. In
Figure \ref{IRXUV_UV_M81_SFH}, for instance, the scenario with
$\tau_\mathrm{SF} = 0.1$ Gyr offers a description of age $<$1 Gyr
for the local background regions, while in the scenario with
$\tau_\mathrm{SF} = 0.5$ Gyr most of these regions correspond to
stellar population age extending to 8 Gyr. Due to the variations in
SFH, it is very likely for stellar populations with a large
disparity in age to lie close to each other or even overlap in the
IRX-UV diagram, and in this case, the appearance of the age
parameter tends to be of less prominence in the IRX-UV function.

The separation between the UV clusters and the local background
regions in the IRX-UV diagram is obviously displayed for NGC~4536
and NGC~7331 (aside from the ring area), visible but not very
apparent to distinguish for NGC~6946, and completely absent for
NGC~5194 (Figures \ref{IRXUV_UV_M51}$-$\ref{IRXUV_UV_NGC7331}). When
the diversity of SFH is applied to these galaxies, the local
background regions are expected to have longer star formation
timescales than the UV clusters, and the factual difference in
stellar population age between the two components inside each galaxy
tends to exceed the interpretation of the offset in the IRX-UV
diagram on assumption of one identical SFH. Combining the data
points in Figures \ref{IRXUV_UV_M51}$-$\ref{IRXUV_UV_NGC7331} and
the modeled curves in Figure \ref{IRXUV_UV_M81_SFH}, we can see
that, most of the local background regions are characterized to
cover an age range of 1$-$8 Gyr with $\tau_\mathrm{SF} = 0.5$ Gyr,
and longer star formation timescales assign the same age range for
bluer color spaces. In such cases, the UV clusters and the local
background regions are able to possess the adjacent or even common
regimes in $\mathrm{FUV}-\mathrm{NUV}$ at similar levels of IRX, in
spite of possible disparities in stellar population age.

However, the variations in SFH still fail to provide any better
solution to the other discrepancies in characterizing the IRX-UV
relation for the galaxies with the scenario of simple stellar
populations presented in Section \ref{result1}. It is evident that, a few number of UV clusters inside NGC~3031 and several local
background regions inside NGC~5194 which have bluer UV colors at
fixed IRX than the prediction of the scenario with an instantaneous
burst remain outside the coverage of the grid modeled with any other
SFH, and a half of UV clusters inside NGC~5194 and ring clusters
inside NGC~7331 which are estimated as several hundreds of
megayears assuming an instantaneous burst possess even
older regimes in age with increasing star formation timescales.

Figure \ref{IRXUV_UV_total} shows significant dispersion in the
composite IRX-UV relation for a total number of the UV clusters, and
the scenario with an instantaneous burst fails to provide an
adequate interpretation of the data distribution. The variations in
SFH are shown in Figure \ref{IRXUV_UV_M81_SFH} to have an
effective impact on the IRX-UV properties for stellar populations
with age $\geq$100 Myr. For instance, between the SFHs with an
instantaneous burst and the $\tau_\mathrm{SF} = 0.1$ Gyr star
formation, there is a scatter of $\sim$0.3 mag in
$\mathrm{FUV}-\mathrm{NUV}$ at fixed IRX for the age of 100 Myr, and
a scatter of $\sim$0.6 mag in $\mathrm{FUV}-\mathrm{NUV}$ for 300
Myr. Older ages and longer star formation timescales are not taken
into consideration for the UV clusters. In this situation, although
UV clusters are believed to contain much simpler stellar populations
than galactic background, the limited variations in SFH are still
likely to introduce a moderate level of dispersion in Figure
\ref{IRXUV_UV_total}, particularly in view of the different physical
scales of the measured regions inside different galaxies (see
Section \ref{measure}).

The adoption of various SFHs complements the interpretation of the
data distributions with the model assuming simple stellar
populations. However, it remains deficient for the current scenarios
to well characterize all the data. As a number of studies have
suggested, various dust grain properties and diverse spatial
geometries are suspected of causing discrepancies in the relationship
between attenuation and observed color indices at certain wavebands,
which thus leads to different IRX-UV trends
\citep[e.g.,][]{2000ApJ...528..799W, 2006MNRAS.370..380I}. In Paper II, we will
investigate the dependence of the IRX-UV relation on
attenuation/extinction law and attempt to discover more parameters
in the IRX-UV function, speciﬁcally in order to obtain solutions from
an alternative viewpoint to the discrepancies between the
observational data and the current scenarios.

Statistical studies of integrated galaxies have shown large
dispersion in the IRX-UV relation
\citep[e.g.,][]{2005ApJ...619L..51B, 2007ApJ...655..863D,
2009ApJ...703..517D, 2007ApJS..173..185G}. Although stellar
population age has been widely considered as the second parameter for
introducing the deviation, until now there has not been sufficient
evidence to confirm the age effect on the IRX-UV relation for
integrated galaxies \citep{2005ApJ...619L..55S,
2007ApJS..173..392J}. In this paper, the comparisons between the
spatially resolved measurements and the integrated properties of
galaxies provide implications about the conundrum in integrated
studies of galaxies. The locations of integrated galaxies and
galactic subregions in the IRX-UV diagrams (Figures
\ref{IRXUV_UV_M81}$-$\ref{IRXUV_UV_NGC7331}) and the fractional
contributions of component to integrated luminosities of galaxies
(Table \ref{contri}) indicate that, any single component lacks the
ability to dominate the integrated luminosities of galaxies, and
integrated measurements of normal galaxies are aggregations of
various populations and different components within galaxies. As
also implied in \citet{2012A&A...539A.145B}, simple age tracers are
not in a good position to estimate stellar population age in such a
case. If galaxies exhibit prominent features in sub-structures at
certain observational bands, the integrated measurements can have an
effective impact. In most cases of integrated measurements which
compromise different stellar populations in galaxies and
consequently represent more complexities in SFH, contrary to
spatially resolved studies, any significant variation in the IRX-UV
trend as a function of stellar population age tends to become
unclear and probably masked by other potential factors. Moreover,
even though in the simple situation any other parameter is
negligible, the effects of stellar population age are suggested to
be more complicated than we have realized in the IRX-UV plane
\citep{2008MNRAS.386.1157C}.

\section{SUMMARY}\label{sum}

In this work, we perform a spatially resolved analysis of five
spiral galaxies selected from the SINGS sample, aimed at investigating
the impacts of dust attenuation and stellar population age on the
IRX-UV relation, and thereby to provide better insights into the
influences of dust and stellar population properties on UV and IR
observations of galaxies.

Aperture photometry is extracted for all positions within the
galaxies. We divide the measured regions into UV clusters and local
background regions. For NGC~3031 we further divide the local
background regions into disk regions and bulge regions, and for
NGC~7331 we mark the regions in the ring area, in order to
illuminate their special properties. The classification of different
populations can help to distinguish the signature of stellar
population age in the IRX-UV function. We also measure the
integrated luminosities of the entire galaxies and the galactic
center areas in order to compare the integrated and spatially
resolved properties of galaxies.

One of our main results is the separation of the effects of age and dust
attenuation on the IRX-UV relation which failed to be
revealed by studies of integrated galaxies
\citep[e.g.,][]{2005ApJ...619L..55S, 2007ApJS..173..392J}. The age
signature is indicated by systematic offsets of the local
background regions towards redder color ranges from the UV clusters
in the IRX-UV diagrams due to the intrinsic redder UV colors for
evolved stellar populations. This kind of displacement is clearly
seen in NGC~3031, NGC~4536, and NGC~7331; distinguishable but less prominent in NGC~6946; and completely invisible in NGC~5194.
When we attempt to constrain variations in the age parameter by
plotting all the UV clusters in one diagram, the composite relation
doesn't display a common trend, and instead presents a considerable
degree of scatter. Variations in SFH are suggested as a potential
cause of weakening the age effects on the IRX-UV relation, and we
therefore interpret the different levels of overlapping between the
UV clusters and the local background regions, as well as a few
portions of the dispersion in the composite relation for all the UV
clusters. However, definite discrepancies still
appear in the descriptions of the data loci for NGC~3031,
NGC~5194, NGC~7331, and the total number of UV clusters by the
scenarios assuming any of the SFHs. These difficulties indicate the
necessity of other parameters such as attenuation/extinction law in
operation of the IRX-UV function. The coincident requirement was
suggested in \citet{2009ApJ...706..553B, 2012A&A...539A.145B}. With
this in consideration, further examinations involving attenuation/extinction law will be under taken in Paper II.

Through the investigations into the integrated properties of
galaxies and the fractional contributions of different components to
the integrated measurements, we find that the UV clusters account
for $\sim20\%-30\%$ of the overall luminosities of galaxies, and the
integrated characters are represented by the local background in
most cases. If galaxies perform prominent features in substructures,
the organizational regime will be reformed, and integrated
measurements of galaxies tend to be biased by the substructures in
this case: in our sample the star-forming center in NGC~4536 and the
dust ring in NGC~7331 play dominant roles in the integrated
measurements of their host galaxies. The subregions in NGC~6946
present a vast scatter in the IRX-UV diagram, and consequently, the location of the integrated galaxy has the largest
deviation from the starburst empirical curve in our sample.

In addition, the IRX-UV relation depends weakly on
luminosity showing that either $\mathrm{FUV}-\mathrm{NUV}$ color or
IRX is inversely related to FUV luminosity, differing from the
results from the statistical studies of integrated galaxies, which is
due to the different natures of the sources sampled between the two
kinds of studies. The radial trends for $\mathrm{FUV}-\mathrm{NUV}$
color and IRX present declining gradients from the center to the edge in
galaxies. From the two-dimensional maps, the different spatial
distributions between the two parameters can be further seen: the
$\mathrm{FUV}-\mathrm{NUV}$ color descends with increasing galactic
radii more symmetrically and smoothly, whereas the IRX exhibits
several sorts of substructures such as spiral arms and clumps. This
discrimination provides a spatial mirror of the dispersion in the
IRX-UV relation.

\acknowledgments

We appreciate a number of substantial contributions from Benjamin
Johnson to this work. We are grateful to the anonymous referee for
insightful and constructive comments which have greatly improved the
paper. This work is supported by the National Natural Science
Foundation of China (NSFC, Nos. 10873012, 10833006 and 11003015), the 
Open Research Program of Key Laboratory for the Structure and 
Evolution of Celestial Objects, CAS, and Chinese Universities 
Scientific Fund (CUSF). Ye-Wei Mao thanks Caitlin Casey, Quinton 
Goddard, and Wei Zhang for their technical assistance at early stages 
of this work. Ye-Wei Mao acknowledges the support of the China 
Scholarship Council. This research has made use of the NASA/IPAC 
Extragalactic Database (NED), which is operated by the Jet Propulsion 
Laboratory, California Institute of Technology, under contract with 
the National Aeronautics and Space Administration. This research has 
also made use of the NASA's Astrophysics Data System.

\appendix

\section{ESTIMATION OF TOTAL IR LUMINOSITY}\label{App}

In this work, the observed total IR luminosity L(IR) is estimated
from 8 and 24 $\mu$m luminosities via C05 calibration (Equation
(\ref{C05_cali}) in this paper), since the poor resolution of
\emph{Spitzer} MIPS 70 and 160 $\mu$m observations hinders us from
applying longer wavelength imaging. C05 calibration takes the
estimation of IR luminosity from 24, 70, and 160 $\mu$m luminosities
in \citet[][Equation (4) therein. Hereafter denoted as
DH02]{2002ApJ...576..159D} as the reference of L(IR). The sample used in
C05 calibration is star-forming regions in NGC~5194, and for these
regions a predominant fraction of the total IR luminosity 
originates from short wavelength components of IR continua. In this
situation, C05 calibration is suitable for young stellar populations
such as the UV clusters in our work. For more evolved populations
such as the local background regions or integrated galaxies, it is not clear whether
or not the same prescription can still provide appropriate
estimations. In consideration of this question, here we compare
total IR luminosities derived from C05 and DH02 methods, and examine
the usability of C05 calibration in more general cases.

\begin{figure}[!h]
\centering
\vspace*{-10mm}
\includegraphics[width=0.8\columnwidth]{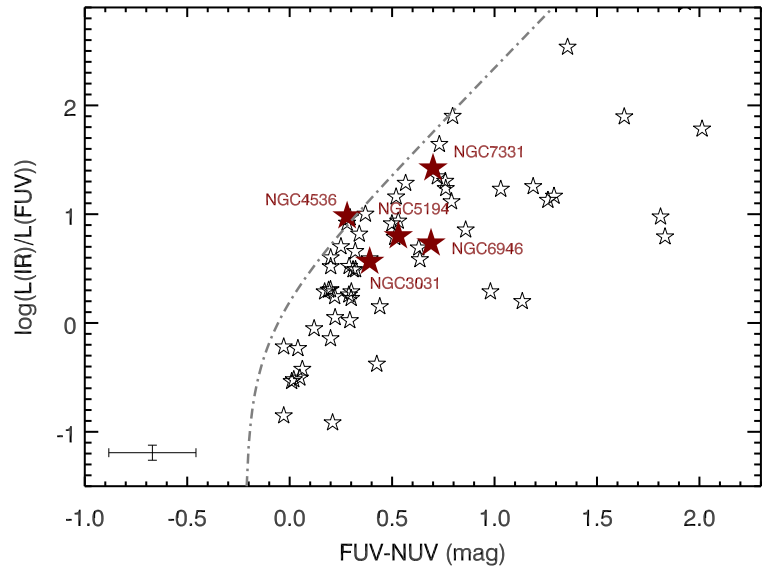}
\vspace*{-85mm}
\caption{The same diagram with Figure \ref{IRXUV_D07} but the total
IR luminosity is estimated by adopting DH02 calibration (i.e., L(IR)
is derived from combination of L(24 $\mu$m), L(70 $\mu$m), and L(160
$\mu$m)).} \label{IRXUV_D07_2}
\end{figure}

\begin{figure}[!h]
\centering
\vspace*{-10mm}
\includegraphics[width=0.8\columnwidth]{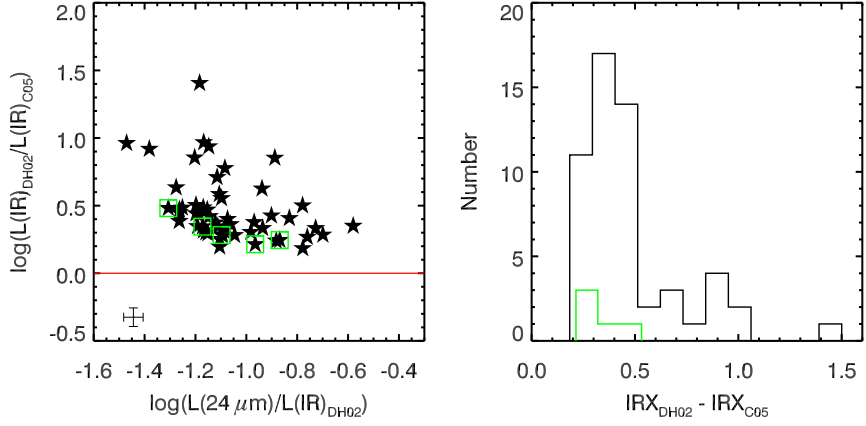}
\vspace*{-90mm}
\caption{Comparison between C05 and
DH02 total IR calibrations for the integrated SINGS galaxies. Left:
DH02-to-C05 L(IR) ratio as a function of L(24
$\mu$m)-to-L(IR)$_\mathrm{DH02}$ ratio. The SINGS galaxies are
symbolized as black filled stars and the galaxies studied in this
paper are enclosed in green boxes. The error bar at the corner shows
the median uncertainty for the sample. Right: Histogram of margins
between DH02-based IRX and C05-based IRX for the SINGS sample (black
line), and the galaxies studied in this paper are also shown (green
line).} \label{compTIR}
\end{figure}

In Figure \ref{IRXUV_D07}, we plot the IRX-UV diagram for the D07
integrated measurements of the SINGS galaxies, where the total IR
luminosity is calculated by employing C05 calibration (8 and 24
$\mu$m). For the purpose of comparison, here we plot the same
diagram in Figure \ref{IRXUV_D07_2} but estimate L(IR) by adopting
DH02 prescription (24, 70, and 160 $\mu$m). The data loci in the
both diagrams hold a consistent degree of scatter, but in Figure
\ref{IRXUV_D07_2} the data points lie at a higher level in the
diagram and populate closer to the starburst empirical line. In
Figure \ref{compTIR}, we further compare C05- and DH02-calibrated
total IR luminosities, L(IR)$_\mathrm{C05}$ (8-24 $\mu$m
calibration) and L(IR)$_\mathrm{DH02}$ (24-70-160 $\mu$m
calibration), for the same sample. The left panel of Figure
\ref{compTIR} shows
$\log(L(\mathrm{IR})_\mathrm{C05}/L(\mathrm{IR})_\mathrm{DH02})$ as
a function of $\log(L(24~\mu
\mathrm{m})/L(\mathrm{IR})_\mathrm{DH02})$. For all data points,
DH02 estimation yields larger L(IR) than C05 prescription, with
margins of 0.2$-$0.5 dex for most galaxies. At the same time, this
diagram shows an obvious trend that, when L(24 $\mu$m) accounts for
a tiny minority of L(IR), there is a vast offset between
L(IR)$_\mathrm{C05}$ and L(IR)$_\mathrm{DH02}$, even as much as 1.0
dex; but if the contribution of L(24 $\mu$m) to L(IR) increases, the
offset tends to become of less significance, and
L(IR)$_\mathrm{C05}$ is therefore comparable to
L(IR)$_\mathrm{DH02}$. The offset in IRX introduced by the different
IR calibrations is apparently shown in the right panel of Figure
\ref{compTIR}. The C05-based IRXs appear to be lower by a
typical value of $\sim$0.4 than DH02 products.

\begin{figure}[!h]
\centering
\vspace*{-10mm}
\includegraphics[width=0.8\columnwidth]{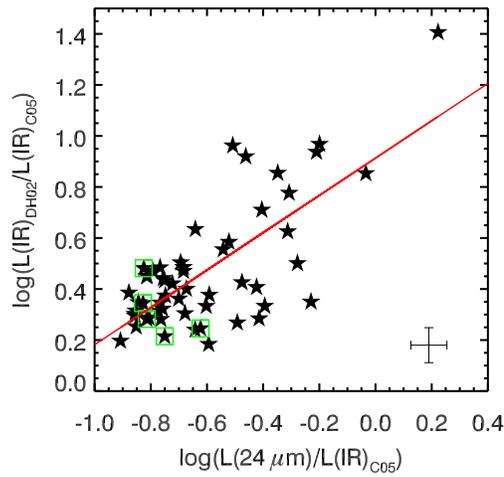}
\vspace*{-90mm}
\caption{DH02-to-C05 L(IR) ratio as a function of L(24
$\mu$m)-to-L(IR)$_\mathrm{C05}$ ratio. Symbols are the same with
those in the left panel of Figure \ref{compTIR}. The red solid line
is the best-fitting curve to the SINGS sample, defined as Equation
(\ref{eq_A1}). The error bar at the corner shows the median
uncertainty for the sample.} \label{cali_TIR}
\end{figure}

\begin{figure}[!h]
\centering
\vspace*{-10mm}
\includegraphics[width=0.8\columnwidth]{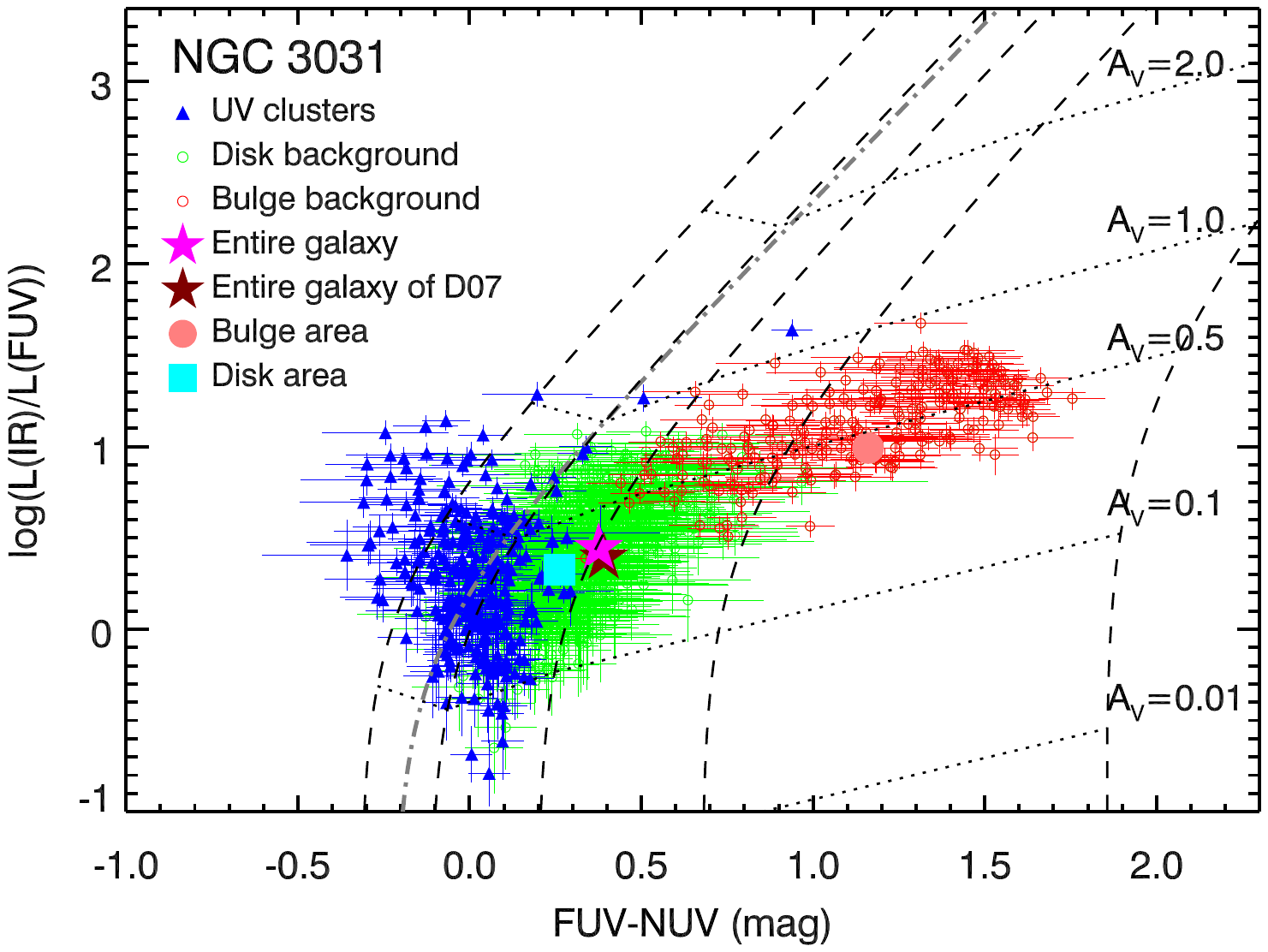}
\vspace*{-85mm}
\caption{IRX vs. $\mathrm{FUV}-\mathrm{NUV}$ for NGC~3031. Symbols
and lines are the same with those in Figure \ref{IRXUV_UV_M81}, but
the total IR luminosity L(IR) is derived from L(24 $\mu$m) and
L(IR)$_\mathrm{C05}$ by adopting Equation (\ref{eq_A1}). Error bars
showing the photometric uncertainties are plotted as well.}
\label{IRXUV2_UV_M81}
\end{figure}

\begin{figure}[!h]
\centering
\vspace*{-10mm}
\includegraphics[width=0.8\columnwidth]{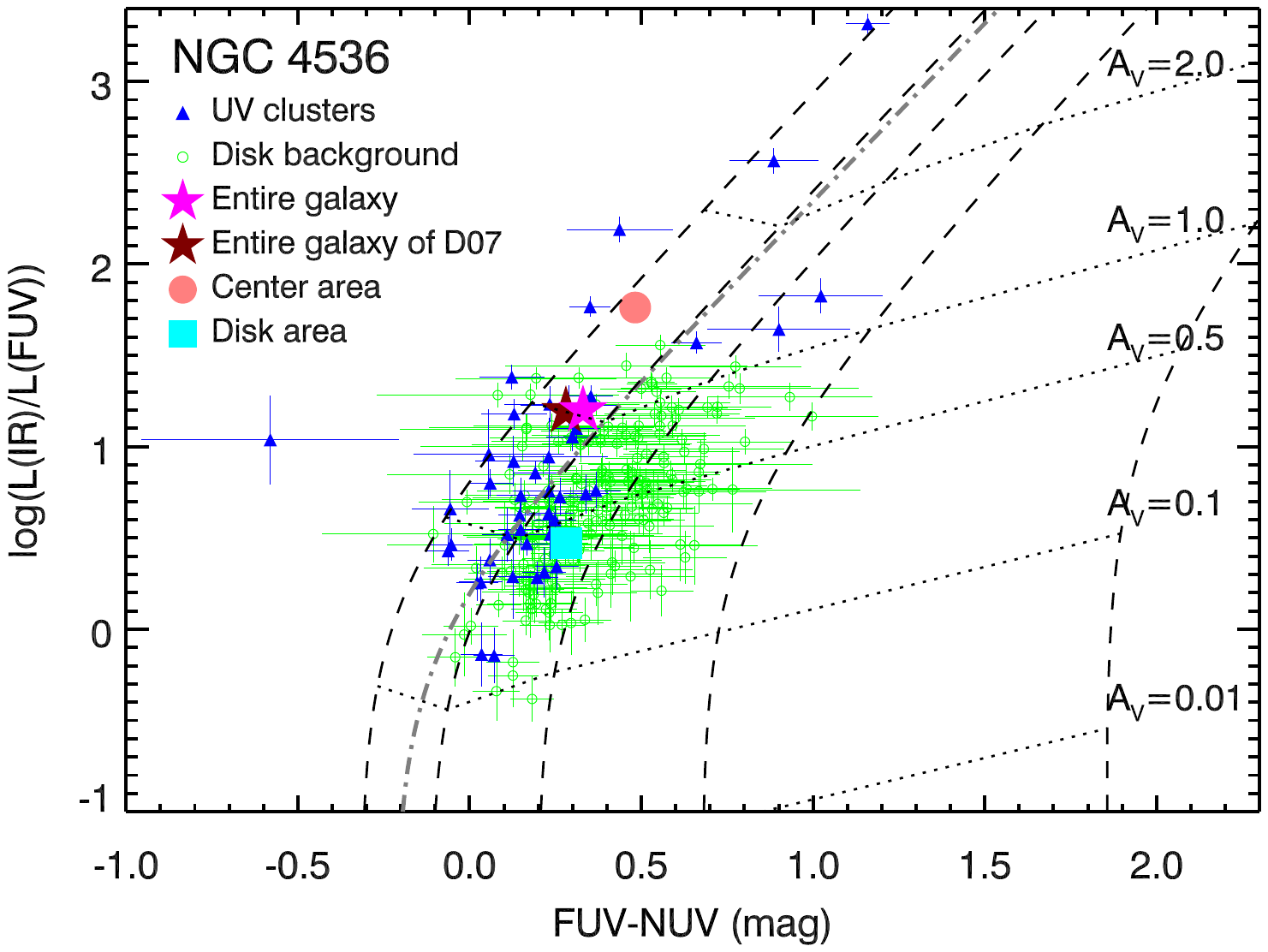}
\vspace*{-85mm}
\caption{IRX vs. $\mathrm{FUV}-\mathrm{NUV}$ for NGC~4536. Symbols
and lines are the same with those in Figure \ref{IRXUV_UV_NGC4536},
but the total IR luminosity L(IR) is derived from L(24 $\mu$m) and
L(IR)$_\mathrm{C05}$ by adopting Equation (\ref{eq_A1}). Error bars
showing the photometric uncertainties are plotted as well.}
\label{IRXUV2_UV_NGC4536}
\end{figure}

\begin{figure}[!h]
\centering
\vspace*{-10mm}
\includegraphics[width=0.8\columnwidth]{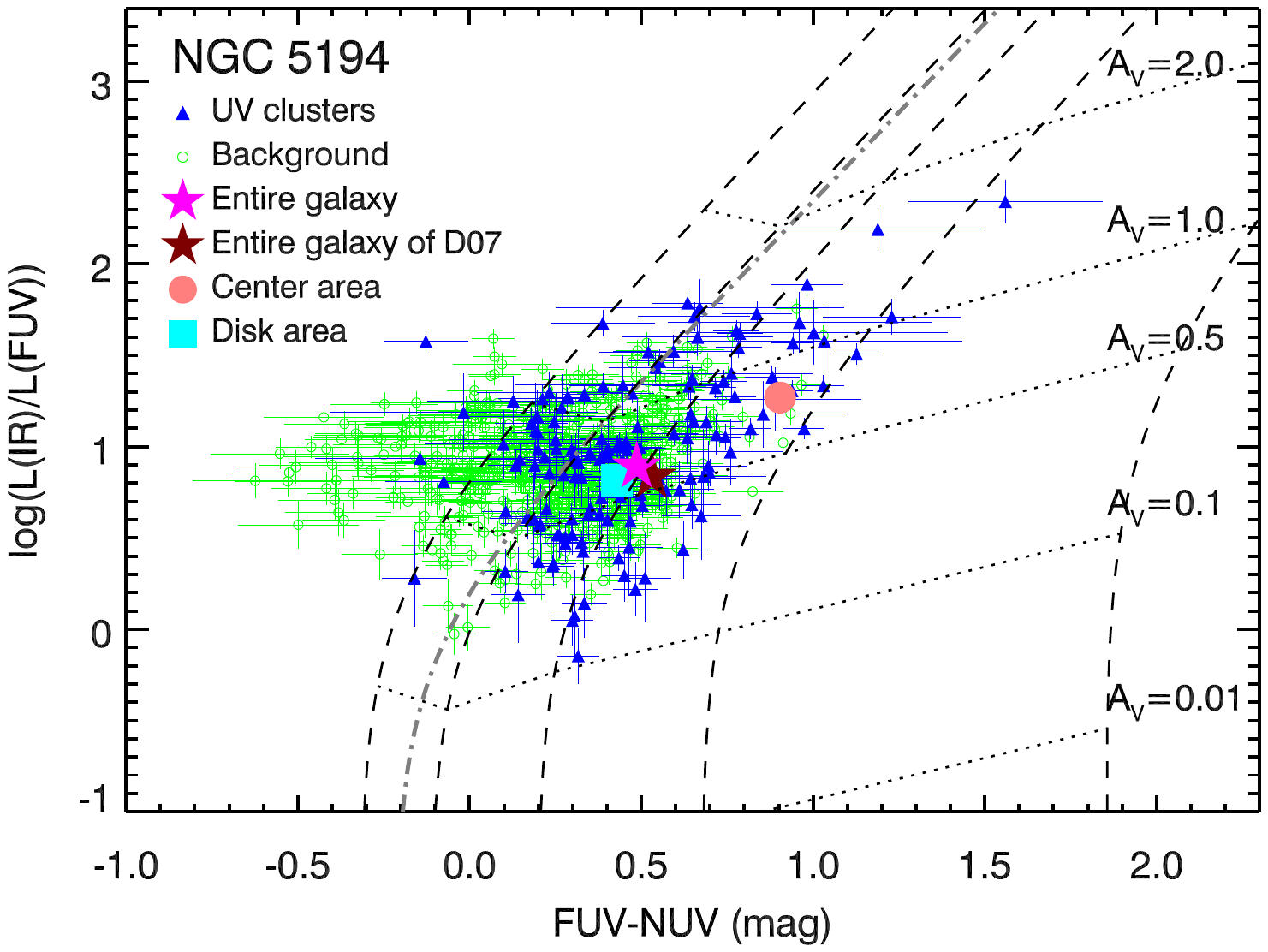}
\vspace*{-85mm}
\caption{IRX vs. $\mathrm{FUV}-\mathrm{NUV}$ for NGC~5194. Symbols
and lines are the same with those in Figure \ref{IRXUV_UV_M51}, but
the total IR luminosity L(IR) is derived from L(24 $\mu$m) and
L(IR)$_\mathrm{C05}$ by adopting Equation (\ref{eq_A1}). Error bars
showing the photometric uncertainties are plotted as well.}
\label{IRXUV2_UV_M51}
\end{figure}

\begin{figure}[!h]
\centering
\vspace*{-10mm}
\includegraphics[width=0.8\columnwidth]{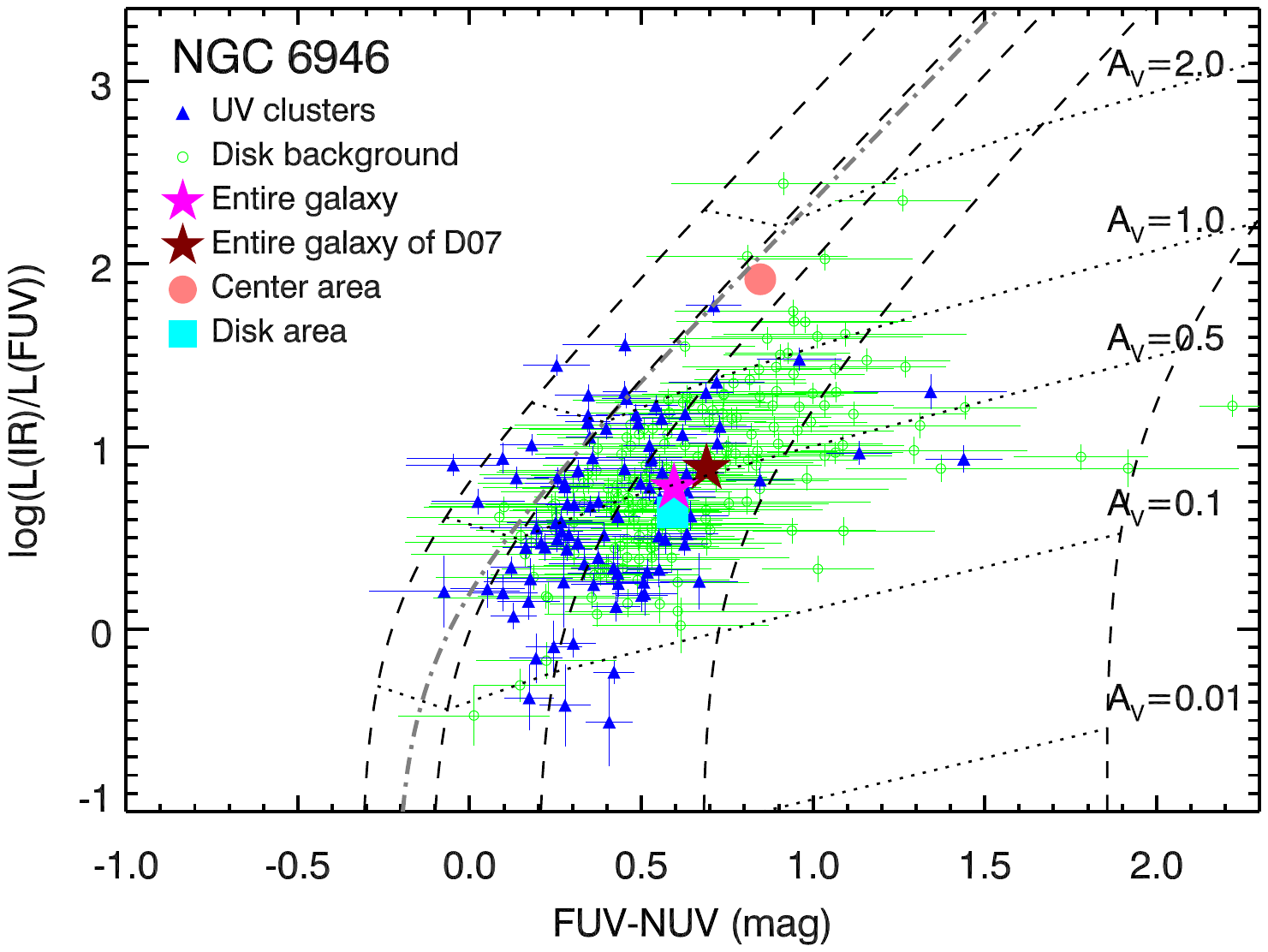}
\vspace*{-85mm}
\caption{IRX vs. $\mathrm{FUV}-\mathrm{NUV}$ for NGC~6946. Symbols
and lines are the same with those in Figure \ref{IRXUV_UV_NGC6946},
but the total IR luminosity L(IR) is derived from L(24 $\mu$m) and
L(IR)$_\mathrm{C05}$ by adopting Equation (\ref{eq_A1}). Error bars
showing the photometric uncertainties are plotted as well.}
\label{IRXUV2_UV_NGC6946}
\end{figure}

\begin{figure}[!h]
\centering
\vspace*{-10mm}
\includegraphics[width=0.8\columnwidth]{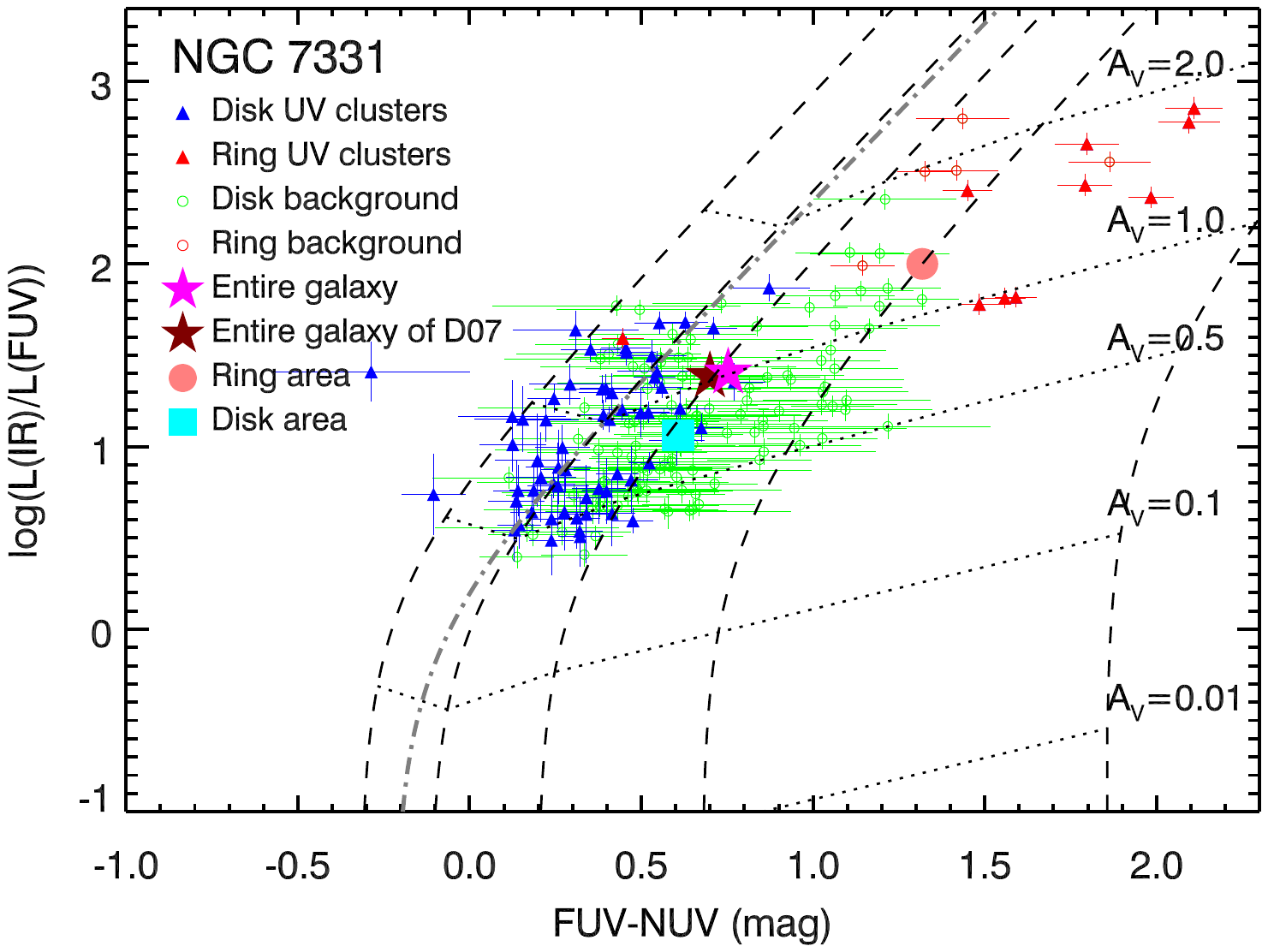}
\vspace*{-85mm}
\caption{IRX vs. $\mathrm{FUV}-\mathrm{NUV}$ for NGC~7331. Symbols
and lines are the same with those in Figure \ref{IRXUV_UV_NGC7331},
but the total IR luminosity L(IR) is derived from L(24 $\mu$m) and
L(IR)$_\mathrm{C05}$ by adopting Equation (\ref{eq_A1}). Error bars
showing the photometric uncertainties are plotted as well.}
\label{IRXUV2_UV_NGC7331}
\end{figure}

\begin{figure}[!h]
\centering
\vspace*{-10mm}
\includegraphics[width=0.8\columnwidth]{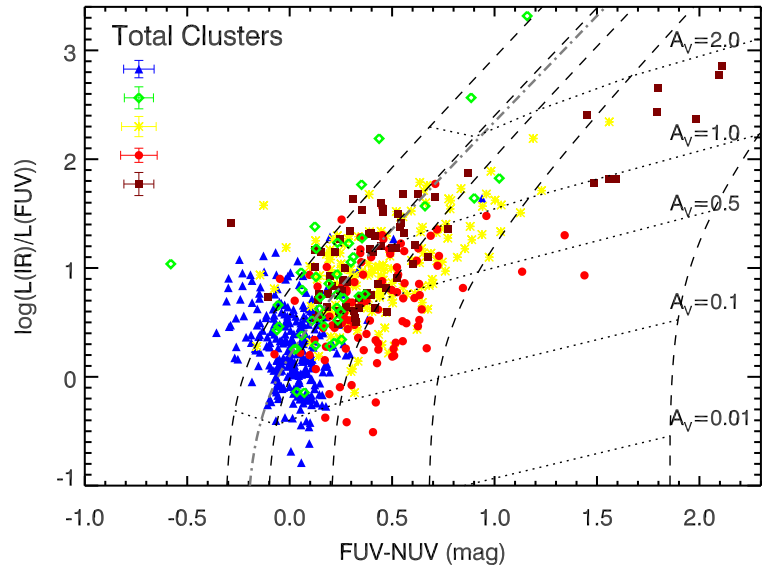}
\vspace*{-85mm}
\caption{IRX vs. $\mathrm{FUV}-\mathrm{NUV}$ for the total UV
clusters inside the five galaxies in our sample. Symbols and lines
are the same with those in Figure \ref{IRXUV_UV_total}, but the
total IR luminosity L(IR) is derived from L(24 $\mu$m) and
L(IR)$_\mathrm{C05}$ by adopting Equation (\ref{eq_A1}). Error bars
showing the median uncertainty for each galaxy are plotted at the
higher left corner.} \label{IRXUV2_UV_total}
\end{figure}

The analysis of IR estimation implies that, the total IR
luminosities derived via C05 calibration in this paper are likely to
be underestimated, particularly for the local background regions and
the integrated galaxies. Figure \ref{cali_TIR} shows the correlation
between
$\log(L(\mathrm{IR})_\mathrm{DH02}/L(\mathrm{IR})_\mathrm{C05})$ vs.
$\log(L(24~\mu \mathrm{m})/L(\mathrm{IR})_\mathrm{C05})$ for the
SINGS galaxies with the D07 measurements, which offers a rough
recipe to estimate L(IR)$_\mathrm{DH02}$ from L(IR)$_\mathrm{C05}$
and L(24 $\mu$m) and therefore to compensate the possible bias in
L(IR)$_\mathrm{C05}$. The best-fitting curve for the data points in
this figure provides the equation:
\begin{equation}
\log(L(\mathrm{IR})_\mathrm{DH02}) =
\log(L(\mathrm{IR})_\mathrm{C05}) + 0.913 + 0.732 \log(L(24~\mu
\mathrm{m})/L(\mathrm{IR})_\mathrm{C05}). \label{eq_A1}
\end{equation}
We adopt this equation to re-estimate the total IR luminosities for
the galaxies in our sample, and the resulting IRX-UV diagrams are
displayed in Figures \ref{IRXUV2_UV_M81}$-$\ref{IRXUV2_UV_total}.
The new version of the IRX-UV relations presents shapes of loci generally
consistent with those in Figures
\ref{IRXUV_UV_M81}$-$\ref{IRXUV_UV_total}, but the data points
systematically shift to higher IRX levels by an average factor of
0.4. The change in total IR luminosity has no impact on the age
signature, and the offset between the UV clusters and the local
background regions remains to be seen for NGC~3031, NGC~4536,
NGC~6946, and NGC~7331, with different levels of overlapping. The discrepancies existing in the distributions for
NGC~3031, NGC~5194, and NGC~7331 are still unsolved in the updated
diagrams, and the significant dispersion remains in the composite
relation for the total UV clusters.

Equation (\ref{eq_A1}) serves as an approximate modification of C05
calibration applied to evolved stellar populations. The direct
purpose of introducing this correction is to examine the potential
deviation in the IR estimation in this paper. It is necessary to
emphasize that, the conversion between L(IR)$_\mathrm{DH02}$ and
L(IR)$_\mathrm{C05}$ is questionable in more general cases. For
young stellar populations such as extragalactic star-forming
regions, C05 calibration is still supposed to be an optimal approach
to assessing total IR luminosity when only 8 and 24 $\mu$m data are
available.

The difference between the results from DH02 and C05 calibrations
arises primarily from different dust temperatures triggered by young
and evolved stellar populations. As an example, for NGC~3031 the
dust heated by diffuse emission of evolved stellar populations has
the temperature of about 19 K and the corresponding IR continuum
peaks at around 160 $\mu$m, whereas young ionizing sources trigger
the dust temperature over 100 K with the consequent IR luminosity
peaking at around 24 $\mu$m \citep{2006ApJ...648..987P}; the average
dust temperature for the integrated NGC~3031 is about 30 K and the
IR continuum peaks at around 100 $\mu$m \citep{1995AJ....110.1115D}.
Therefore, for galactic background or integrated galaxies where
evolved populations dominate stellar contents, the cold dust
contributes IR emission, and in this case the estimation of IR
luminosity from short waveband monochromatic fluxes calibrated on a
basis of warm-dust-dominating sources fails to reach the longer
wavelength component of IR continuum and introduces the
underestimate of IR luminosity. The left panel of Figure
\ref{compTIR} suggests a trend that, when warmer dust contributes IR
emission which is reflected by the increasing proportion of 24
$\mu$m to total IR luminosity, the two estimators tend to match with
each other; By contrast, the contribution of colder dust would
augment the deviation. The advent of the \emph{Herschel}
observations with high resolution at FIR wavebands will provide
perceptive insight into cold dust in galaxies and is expected to
fill the gap in longer wavelength components of galactic IR emission
\citep{2011PASP..123.1347K}.

\end{document}